\pdfoutput=1
\documentclass[11pt]{article}
\usepackage{tabularx}
\usepackage[final]{acl}

\usepackage{times}
\usepackage{latexsym}
\usepackage{cuted}
\usepackage{afterpage}

\usepackage{minitoc}
\usepackage{titletoc}
\usepackage{longtable}
\usepackage{tabularx}
\usepackage[T1]{fontenc}
\usepackage[utf8]{inputenc}

\usepackage{microtype}

\usepackage{inconsolata}

\usepackage{bm}
\usepackage{tikzducks}
\usepackage{graphicx}
\usepackage[most]{tcolorbox}
\usepackage{colortbl}
\usepackage{pifont}
\usepackage[table,x11names,dvipsnames]{xcolor}
\usepackage{enumitem}
\definecolor{Gray}{gray}{0.9}

\usepackage{subfigure}
\usepackage{tikz}
\newcommand*\circled[1]{\tikz[baseline=(char.base)]{
            \node[shape=circle,draw,inner sep=0.4pt] (char) {#1};}}

\usepackage{listings}

\lstdefinestyle{mystyle}{
    backgroundcolor=\color{white},
    basicstyle=\fontsize{7.5pt}{7.5pt}\ttfamily\selectfont,
    columns=fullflexible,
    breaklines=true,
    captionpos=b,
    commentstyle=\fontsize{7.5pt}{7.5pt}\color{codeblue},
    keywordstyle=\fontsize{7.5pt}{7.5pt}\color{codekw},
}
\usepackage{graphicx}
\usepackage{amsmath, amssymb, bm}
\usepackage[ruled,vlined]{algorithm2e} 
\usepackage{algpseudocode}
\usepackage{multirow}
\usepackage{graphicx}
\usepackage{booktabs}
\usepackage{multicol}
\usepackage{hyperref}
\usepackage{amsmath}
\usepackage{caption}
\usepackage{enumitem}
\usepackage{wrapfig}
\usepackage{listings}
\usepackage{pifont}
\usepackage{array}
\usepackage{placeins}
\usepackage{float}
\definecolor{codegreen}{rgb}{0.0, 0.411, 0.243}
\definecolor{codered}{HTML}{DB4F59}
\usepackage{hyperref}
\definecolor{dartgreen}{HTML}{00693e}
\definecolor{refcolor}{HTML}{9F363A}
\definecolor{purple}{HTML}{999AC9}

\hypersetup{
    colorlinks=true,
    linkcolor=Bittersweet,
    citecolor=dartgreen,
    filecolor=magenta,      
    urlcolor=dartgreen,
    }

\title{Music Audio-Visual Question Answering Requires Specialized Multimodal Designs}
\author{
 \textbf{Wenhao You}$^{*1}$,
 \textbf{Xingjian Diao}$^{*2}$,
 \textbf{Wenjun Huang}$^{3}$,
 \textbf{Chunhui Zhang}$^{2}$,
 \textbf{Keyi Kong}$^{2}$,
 \\
 \textbf{Weiyi Wu}$^{2}$\textbf{,}
 \textbf{Chiyu Ma}$^{2}$\textbf{,}
 \textbf{Zhongyu Ouyang}$^{2}$\textbf{,}
 \textbf{Tingxuan Wu}$^{4}$\textbf{,}
 \textbf{Ming Cheng}$^{2}$\textbf{,}
   \\
 \textbf{Soroush Vosoughi}$^{2}$\textbf{,}
 \textbf{Jiang Gui}$^{2}$
\\
 $^{1}$University of Waterloo, 
 $^{2}$Dartmouth College,
 $^{3}$UC Irvine,
 $^{4}$New York University
 \\
  \texttt{w22you@uwaterloo.ca}, 
   \texttt{xingjian.diao.gr@dartmouth.edu}
}
\begin{document}
\maketitle

\newcommand\blfootnote[1]{%
  \begingroup
  \renewcommand\thefootnote{}\footnote{#1}%
  \addtocounter{footnote}{-1}%
  \endgroup
}

\blfootnote{$^*$WY and XD contributed equally and are listed alphabetically by their first name. WY, from University of Waterloo, was mentored by XD from Dartmouth College.}

\begin{abstract}
While recent Multimodal Large Language Models exhibit impressive capabilities for general multimodal tasks, specialized domains like music necessitate tailored approaches. Music Audio-Visual Question Answering (Music AVQA) particularly underscores this, presenting unique challenges with its continuous, densely layered audio-visual content, intricate temporal dynamics, and the critical need for domain-specific knowledge. Through a systematic analysis of Music AVQA datasets and methods, this paper identifies that specialized input processing, architectures incorporating dedicated spatial-temporal designs, and music-specific modeling strategies are critical for success in this domain. Our study provides valuable insights for researchers by highlighting effective design patterns empirically linked to strong performance, proposing concrete future directions for incorporating musical priors, and aiming to establish a robust foundation for advancing multimodal musical understanding. We aim to encourage further research in this area and provide a GitHub repository of relevant works: \url{https://github.com/WenhaoYou1/Survey4MusicAVQA}.

\end{abstract}

\begin{figure}[tb]
\centering
\resizebox{0.48\textwidth}{!}{
\includegraphics{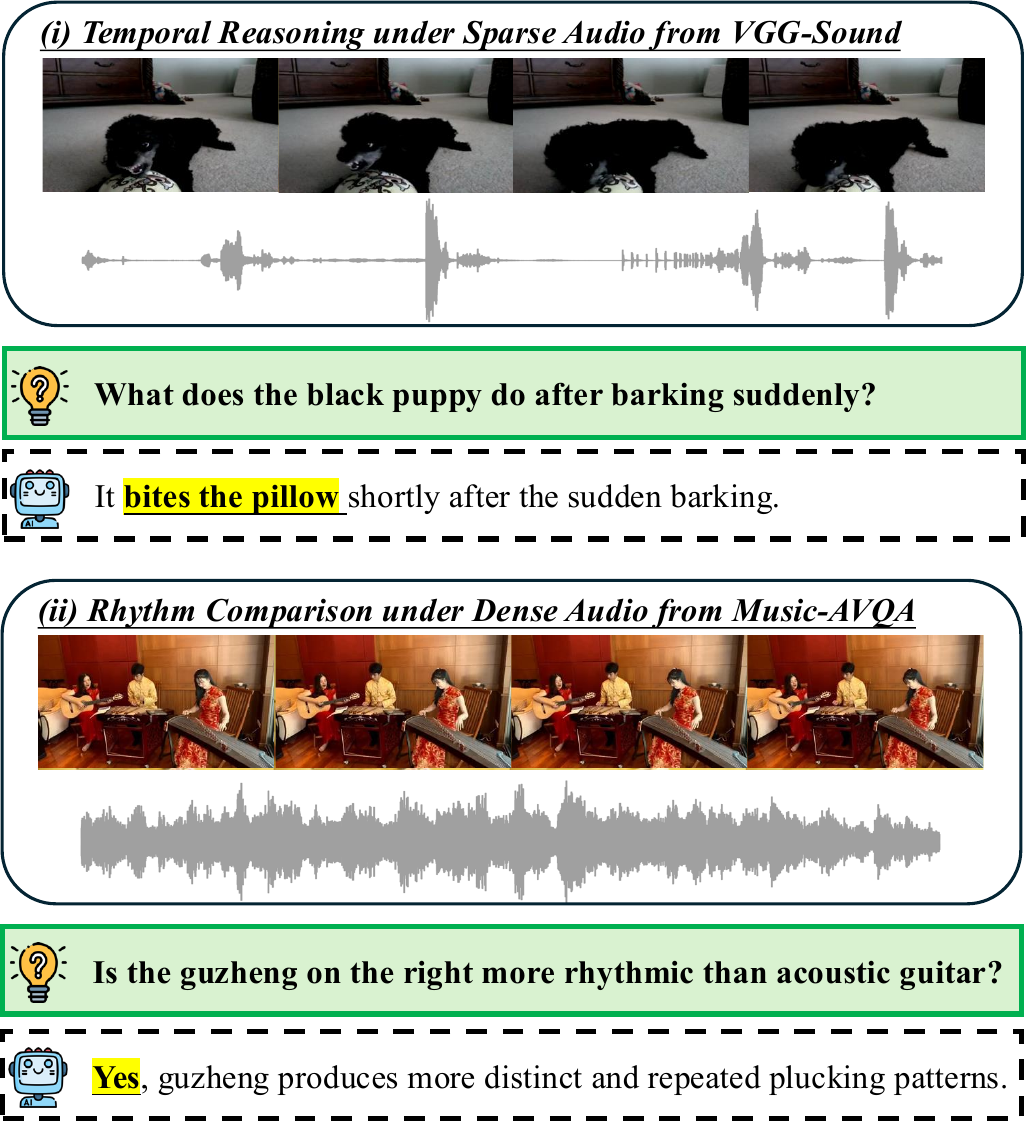}
}
\caption{Contrast between (i) conventional QA and (ii) Music-AVQA with dense audio. Panel (i) shows an isolated sound (barking) and synchronized action, which are relatively easy to detect. Panel (ii) exemplifies music's complexity, featuring overlapping instruments and rhythmic patterns. Such dense and continuous audio-visual signals demand fine-grained temporal and spatial reasoning through cross-modal comparisons and they are more challenging than conventional multimodal QA.}
\vspace{-0.3cm}
\label{fig:teaser}
\end{figure}

\section{Introduction}
\begin{quote}
    \textit{``Music is a moral law. It gives a soul to the Universe, wings to the mind, flight to the imagination, a charm to sadness, gaiety and life to everything.''}  
    \begin{flushright}
        --- Plato (c.\ 427--347~BCE, Ancient Greece)
    \end{flushright}
\end{quote}
Multimodal Large Language Models (MLLMs) have demonstrated impressive efficacy across a wide range of tasks, modelling various modalities such as text, image, and audio~\cite{girdhar2023imagebind,han2024onellm,bai2025qwen2,hurst2024gpt,liang2024querying, huang2024audiogpt,diao2025temporal}. 
However, this success brings forth a critical consideration: the tension between the broad applicability of general multimodal approaches and the requirements of specialized domains~\cite{8665366,weck2024muchomusic,liu2024music}. This leads to a central question: Are general-purpose MLLMs, despite their advancements, truly sufficient for all multimodal tasks, especially those demanding deep, domain-specific understanding?

Music Performance Audio-Visual Question Answering (Music AVQA) emerges as a particularly challenging multimodal domain that compellingly illustrates this tension~\cite{9879157, 10484352, jiang-yin-2023-target,ye2024cat}. Music, in its rich complexity, often requires specialized treatment beyond the capabilities of generic models~\cite{le2025natural,bogdanov2019mtg, chu2023qwen, diao2025learning, deng2023musilingo, yuan2024chatmusician, zhao2024openmu, agostinelli2023musiclm, lu2023musecoco}. Unlike common scenarios with sparse and discrete audio signals, music performances exhibit a continuous and tightly interwoven blend of audio and visual signals, offering a uniquely rich context for fine-grained audio-visual scene understanding and temporal reasoning~\cite{9879157, 10484352, diao2024learning, ma2024look, li2024object, li2023progressive}. For instance, tasks such as discerning the loudest instrument amidst an ensemble or comparing rhythmic complexity between two spatially distinct performers demand a level of granularity that general-purpose MLLMs may not inherently possess~\cite{liu2024music,diao2024learning,li2024object}. Thus, Music AVQA serves as an ideal lens through which to examine the limitations of general multimodal approaches and to advocate for the necessity of domain-specific adaptations~\cite{9879157,diao2024learning}.

The unique challenges of Music AVQA, illustrated in Figure~\ref{fig:teaser}, stem from the need to reason over continuous, temporally evolving, and densely layered audio-visual signals. Specific complexities include: \textit{\underline{First,}} musical pieces often contain dense and layered audio information. Multiple overlapping instrumental sources are common, which necessitates fine-grained processing to disentangle and interpret complex auditory scenes. \textit{\underline{Second,}} effective understanding depends on precise temporal alignment. It is crucial to accurately associate visual cues—such as a musician’s actions—with their corresponding auditory outputs. This alignment must occur across multiple timescales and often involves intricate temporal dynamics. \textit{\underline{Third,}} the domain frequently requires specialized knowledge. This includes instrument recognition, familiarity with musical theory (such as rhythm and harmony), and an understanding of performance conventions, whether these are explicit or implicit. \textit{\underline{Finally,}} Music AVQA questions often involve complex spatial-temporal relationships. For example, one may need to track dynamic intensity across simultaneous sources (“Which instrument produces the loudest sound?”) or reason about spatial and temporal rhythmic patterns (“Is the cello on the right more rhythmic than the cello on the left?”). Collectively, these factors underscore the unique and demanding nature of reasoning in Music AVQA.

\textbf{This study argues that Music AVQA is a fundamentally distinct multimodal reasoning task, for which specialized multimodal designs are essential and empirically linked to strong model performance.} As the \textbf{\textit{first}} comprehensive survey in this area, we specifically analyze how tailored designs—particularly in input processing and spatial-temporal architecture—enable more effective music understanding compared to generic multimodal systems. Furthermore, we outline how such specialized approaches, by incorporating deeper musical priors, can further advance the field.

\section{Background}
\label{sec:background}
\paragraph{Why Music AVQA is more challenging than normal multimodal understanding?}
Music AVQA presents several distinctive challenges:
\circled{1} \texttt{Dense Signal Interpretation}: Unlike sparse audio events in conventional AVQA, music performances feature continuous, overlapping instrumental sources that require sophisticated separation and attribution;
\circled{2} \texttt{Hierarchical Temporal Reasoning}: Musical information unfolds across multiple time scales (beats, phrases, sections), demanding models capable of reasoning across these hierarchical structures;
\circled{3} \texttt{Cross-Modal Correspondence}: Establishing reliable associations between visual instrumental actions and their acoustic outputs is complicated by temporal misalignments between physical gestures and the resulting sounds;
\circled{4} \texttt{Domain-Specific Knowledge}: Effective reasoning often depends on implicit musical knowledge, such as instrumental techniques, ensemble conventions, and acoustic properties;
\circled{5} \texttt{Abstract Attribute Quantification}: Questions involving subjective qualities such as ``rhythmic", ``melodic,'' or ``harmonious'' require computational strategies to map linguistic descriptors onto measurable signal properties;
\circled{6} \texttt{Data Scarcity}: The specialized nature of musical performances results in smaller and less diverse datasets compared to general AVQA tasks, limiting the generalization capabilities of trained models.

\paragraph{What are common music performance scene types?} \circled{1} \texttt{Solo Performance} – A single musician showcasing technical skills and artistic expression on one instrument.
\circled{2} \texttt{Ensemble of the Same Instrument} – Multiple musicians playing identical or related instruments, creating unified harmonies and textures.
\circled{3} \texttt{Ensemble of Different Instruments} – Musicians performing with a variety of instruments, producing diverse tonal colors and complex musical interactions.
\circled{4} \texttt{Culture-Specific Ensemble} – Traditional instrumental groups that embody the musical heritage and regional styles of specific cultures. See Appendix Section~\ref{appendix:example-scene-types} for detailed examples.

\paragraph{What are common question types in Music AVQA?}
\circled{1} \texttt{Existential Questions}: Determine whether a sound corresponds to a visible object in the scene (e.g., ``Is this sound from the instrument in the video?").
\circled{2} \texttt{Counting Questions}: Quantify audio-visual elements that require cross-modal integration (e.g., ``How many instruments are sounding in the video?").
\circled{3} \texttt{Location Questions}: Identify the spatial position of sound sources within the visual scene (e.g., ``Where is the first sounding instrument?").
\circled{4} \texttt{Comparative Questions}: Compare properties across different audio-visual elements (e.g., ``Is the instrument on the left louder than the one on the right?").
\circled{5} \texttt{Temporal Questions}: Reason about the timing and sequential relationships between auditory and visual events (e.g., ``Which instrument produces sound before the piano?"). See Appendix Section~\ref{appendix:avqaTypeEg} for detailed examples.

\section{Evolution of MUSIC-AVQA Datasets}
\label{sec:dataset_evolution}
The development of Music AVQA research has been driven by progressively refined datasets addressing specific limitations. As summarized in Table \ref{table:dataset} in Appendix Section \ref{sec:music_dataset}, this evolution began with the \circled{1} \textbf{MUSIC-AVQA} dataset~\cite{9879157}, the first large-scale benchmark designed specifically for AVQA in musical contexts, comprising 9,288 performance videos and 45,867 question-answer pairs across diverse reasoning tasks.
Subsequent research reveal challenges related to data bias and imbalanced answer distributions, prompting the creation of \circled{2} \textbf{MUSIC-AVQA v2.0}~\cite{10484352}, which expands to 10,518 videos and approximately 54,000 question-answer pairs. This version balance 15 biased templates by ensuring no dominant answers exceed 60\% for binary questions or 50\% for multi-class questions, particularly enhancing representation in various question categories.
Building on these foundations, \circled{3} \textbf{MUSIC-AVQA-R}~\cite{ma2024look} introduces robustness evaluation through question rephrasing, expanding the test set from 9,129 to 211,572 questions. With a vocabulary five times larger than the original dataset, MUSIC-AVQA-R distinguishes between head (common) and tail (rare) samples, enabling assessment of model performance in both in-distribution and out-of-distribution scenarios. This progressive refinement of datasets has laid a solid foundation for advancing multimodal understanding and robust evaluation in music performance.

\setlength{\tabcolsep}{5.5pt}
\begin{table*}[htbp]
    \centering
% \footnotesize
    % \scriptsize
    \tiny
    \begin{tabular}{l|>{\centering\arraybackslash}c>{\centering\arraybackslash}c>{\centering\arraybackslash}c>{\centering\arraybackslash}c>{\centering\arraybackslash}c}
    \toprule
    \textbf{\textsc{Method}}  & \textbf{Text Encoder} & \textbf{Visual Encoder} & \textbf{Audio Encoder} & \textbf{S-T}\\ 
    \midrule
  \textsc{Amuse}~\cite{diao2024learning}& Transformer~\cite{NIPS2017_3f5ee243}& Swin-Transformer-v2~\cite{liu2022swin}& HTS-AT~\cite{chen2022hts}& \ding{51}\\
\textsc{Audio Flamingo}~\cite{10.5555/3692070.3693076}& OPT-IML-MAX-1.3B~\cite{iyer2023optimlscalinglanguagemodel}& -& ClapCap~\cite{10448504}& \ding{51}\\
\textsc{AVMoE}~\cite{cheng2025mixtures}& -& Swin-Transformer-v2~\cite{liu2022swin}& HTS-AT~\cite{chen2022hts}& $\times$\\
\textsc{AVSD}~\cite{schwartz2019simple}& LSTM& LSTM& LSTM& $\times$\\
\textsc{AVSiam}~\cite{lin2024siamese}& -& ViT~\cite{radford2021learning}& ViT~\cite{radford2021learning}& $\times$\\
\textsc{AVST}~\cite{9879157}& LSTM& ResNet-18~\cite{he2016deep}& VGGish~\cite{45857}& \ding{51}\\
\textsc{CAT}~\cite{ye2024cat}& LLaMA2-7B~\cite{touvron2023llama}& ImageBind~\cite{girdhar2023imagebind}& ImageBind~\cite{girdhar2023imagebind}& $\times$\\
\textsc{ChatBridge}~\cite{zhao2023chatbridge}& Vicuna-13B~\cite{vicuna2023}& ViT-G~\cite{sun2023evaclipimprovedtrainingtechniques}& BEATs~\cite{10.5555/3618408.3618611}& $\times$\\
\textsc{CIGN}~\cite{mo2023class}& -& ResNet-18~\cite{he2016deep}& ResNet-18~\cite{he2016deep}& \ding{51}\\
\textsc{COCA}~\cite{lao2023coca}& Word Embedding& ResNet-18~\cite{he2016deep}& VGGish~\cite{45857}& $\times$\\
\textsc{CONVLSTM}~\cite{9145807}& LSTM& -& Conv& $\times$\\
\textsc{CrossMAE}~\cite{guo2024crossmae}& -& MAE~\cite{9879206}& AudioMAE~\cite{10.5555/3600270.3602351}& $\times$\\
\textsc{DCL}~\cite{lv2023disentangled}& DeBERTa-V3-Large~\cite{he2021debertav3}& ViT~\cite{radford2021learning}& AST~\cite{gong2021ast}& \ding{51}\\
\textsc{DG-SCT}~\cite{duan2023cross}& -& ViT~\cite{radford2021learning}& HTS-AT~\cite{chen2022hts}& \ding{51}\\
\textsc{EEMC}~\cite{10.1007/978-3-031-72904-1_12}& RoBERTa~\cite{zhuang-etal-2021-robustly}& ViT~\cite{radford2021learning}& VGGish~\cite{45857}& \ding{51}\\
\textsc{FCNLSTM}~\cite{9145807}& LSTM& -& Conv& $\times$\\
\textsc{GPT-4o}~\cite{hurst2024gpt}& Transformer& CLIP-ViT& Transformer& $\times$\\
\textsc{GRU}~\cite{7410636}& LSTM& VGGNet~\cite{simonyan2014very}& -& $\times$\\
\textsc{HCRN}~\cite{le2020hierarchical}& BiLSTM& ResNet-18~\cite{he2016deep}& -& $\times$\\
\textsc{LAST-Att}~\cite{10484352}& LSTM& Swin-Transformer-v2~\cite{liu2022swin}& Audio-Spectrogram-Transformer& \ding{51}\\
\textsc{LAVisH}~\cite{lin2023vision}& -& ViT~\cite{radford2021learning}& ViT~\cite{radford2021learning}& \ding{51}\\
\textsc{LAViT}~\cite{yun2021pano}& Transformer~\cite{NIPS2017_3f5ee243}& Transformer~\cite{NIPS2017_3f5ee243}& Transformer~\cite{NIPS2017_3f5ee243}& \ding{51}\\
\textsc{LSTTA}~\cite{liu2023parameter}& CLIP~\cite{radford2021learning}& CLIP~\cite{radford2021learning}& w2v-Conformer~\cite{gulati2020conformer}& \ding{51}\\
\textsc{MAVEN}~\cite{ma2025fortisavqa}& Mixtral& InternViT-300M-448px~\cite{chen2024internvl}& Transformer& $\times$\\
\textsc{MCAN}~\cite{8953581}& GloVe~\cite{pennington-etal-2014-glove}+LSTM& Faster R-CNN~\cite{NIPS2015_14bfa6bb}& -& $\times$\\
\textsc{MCCD}~\cite{ma2024look}& -& -& -& \ding{51}\\
\textsc{Meerkat}~\cite{10.1007/978-3-031-73039-9_4}& LLaMA2-7B~\cite{touvron2023llama}& CLIP-ViT& CLAP~\cite{10095889}& \ding{51}\\
\textsc{OGM}~\cite{wei2024fly}& -& ResNet-18~\cite{he2016deep}& ResNet-18~\cite{he2016deep}& $\times$\\
\textsc{OneLLM}~\cite{han2024onellm}& LLaMA2-7B~\cite{touvron2023llama}& CLIP-ViT& Unified Multimodal Encoder& $\times$\\
\textsc{OPM}~\cite{wei2024fly}& -& ResNet-18~\cite{he2016deep}& ResNet-18~\cite{he2016deep}& $\times$\\
\textsc{PSAC}~\cite{10.1609/aaai.v33i01.33018658}& Word Embedding& CNN& -& $\times$\\
\textsc{PSTP-Net}~\cite{li2023progressive}& CLIP~\cite{radford2021learning}& CLIP~\cite{radford2021learning}& VGGish~\cite{45857}& \ding{51}\\
\textsc{QaP}~\cite{liang2024querying}& DeBERTa-V2-XLarge& CLIP~\cite{radford2021learning}& CLAP~\cite{10095889}& $\times$\\
\textsc{Qwen2.5-VL}~\cite{bai2025qwen2}& MRoPE~\cite{bai2025qwen2}& ViT~\cite{radford2021learning}& -& $\times$\\
\textsc{RefAtomNet}~\cite{peng2024referring}& BERT& ViT~\cite{radford2021learning}& -& \ding{51}\\
\textsc{VALOR}~\cite{10.1109/TPAMI.2024.3479776}& BERT& CLIP~\cite{radford2021learning}& AST~\cite{gong2021ast}& $\times$\\
\textsc{VAST}~\cite{10.5555/3666122.3669307}& BERT~\cite{devlin2019bert}& ViT~\cite{dosovitskiy2020image}& BEATs~\cite{10.5555/3618408.3618611}& $\times$\\
\textsc{VideoLLaMA-2}~\cite{cheng2024videollama}& Transformer& CLIP~\cite{radford2021learning}& BEATs~\cite{10.5555/3618408.3618611}& \ding{51}\\
\textsc{VITA}~\cite{fu2024vita}& Mixtral~\cite{jiang2024mixtral}& InternViT-300M-448px~\cite{chen2024internvl}& CNN& $\times$\\
\bottomrule
\end{tabular}
\caption{
Architectural summary of representative Music AVQA methods. 
Each method lists the text, visual, and audio encoders used, along with an indication of whether explicit spatial-temporal (S-T) modeling is incorporated. 
Detailed descriptions of each method are provided in Appendix~\ref{baseline} and~\ref{sec:existing_methods}.}
\label{tab:methods}
\end{table*}

\section{Categorization of Music AVQA Methods Based on Architecture}
\label{sec:categorization}

Music AVQA methods exhibit diverse architectural designs, particularly in how they encode and integrate textual, visual, and auditory modalities. To better organize existing approaches by their core modeling strategies, we categorize them into three groups—Transformer-based, CNN-based, and Hybrid models—as summarized in Table~\ref{tab:methods}. This categorization highlights how different models are structured to handle the continuous and densely layered nature of musical performances.

\noindent\textbf{Transformer-based models.} Transformer-based models are characterized by the extensive use of self-attention mechanisms, which benefit in particular from their ability to handle long-range temporal dependencies and fine-grained cross-modal alignment. Methods such as Amuse utilize transformers across all modalities, combining a Swin Transformer for visual processing with an HTS-AT transformer for audio encoding, and employing cross-modal adapters to facilitate early and frequent fusion of multimodal information. Similarly, LAST-Att integrates a Swin-V2 Transformer for vision and an Audio Spectrogram Transformer (AST) for audio, emphasizing fine-grained spatial-temporal alignment through pixel-level cross-modal attention. Other methods such as LAVisH and LSTTA, adopt lightweight transformer adapters to inject multimodal cues into frozen transformer backbones, enabling efficient cross-modal reasoning while leveraging strong pre-trained representations.

\noindent\textbf{CNN-based models.} CNN-based methods utilize convolutional backbones such as ResNet or VGGish to encode modality-specific information into global or regional features, often relying on simpler late-stage fusion strategies. The AVST method exemplifies this approach, combining ResNet-18 visual embeddings and VGGish audio features through spatial attention modules to explicitly localize sound sources within visual frames. PSTP-Net extends this design by introducing a progressive refinement strategy that sequentially filters temporal segments and spatial regions, systematically narrowing down question-relevant audio-visual content prior to fusion. Although CNN-based models are computationally efficient and straightforward, their reliance on late fusion may pose challenges to capturing the complex temporal dynamics characteristic of musical performances.

\noindent\textbf{Hybrid models.} Hybrid models combine CNNs, transformers, and large language models (LLMs) to enable unified multimodal reasoning. They typically employ pre-trained encoders from both CNN and transformer families, integrated through sophisticated cross-modal fusion mechanisms. Representative examples include ChatBridge, CAT, OneLLM, and Meerkat. ChatBridge utilizes a perceiver-based multimodal transformer to merge modalities via language-aligned latent representations, followed by a frozen LLM for reasoning. CAT introduces modality-specific clue aggregation modules on top of ImageBind encodings, enabling precise question-driven multimodal grounding before passing information to a generative LLaMA2 LLM. OneLLM further generalizes multimodal integration by introducing a universal projection mechanism that allows a single LLM to interpret diverse modality embeddings seamlessly.

\section{A Call on \textit{Specialized Multimodal Input Processing} for Music AVQA}
\label{sec:processing} 

While input preparation is often treated as a fixed pipeline in general AVQA, music performance settings introduce unique challenges that make input fidelity, segmentation, and representation design especially consequential. Musical scenes are densely layered, temporally continuous, and rich in expressive detail, requiring greater care in how audio, visual, and textual inputs are captured and structured. In what follows, we examine how Music AVQA tasks motivate specialized input processing across three key fronts: maintaining high-resolution and synchronized multimodal signals, adapting tokenization to the structure of musical content, and managing the scale and diversity of music-specific data representations.

\paragraph{Continuous, high-fidelity, and tightly aligned inputs are foundational.}
Compared to event-centric AVQA tasks that typically involve short, discrete sound events and lower-resolution recordings, Music AVQA deals with continuous, polyphonic streams spanning multiple spatial and temporal scales. Audio is commonly sampled at high rates (44.1 kHz or above) and often preserved in lossless formats to retain subtle timbral and articulatory detail~\cite{slizovskaia2021conditioned}. Visual inputs tend to require higher resolution (1080p or above) and frame rates (30–60 fps) to capture nuanced performer motions such as bowing or fingering~\cite{gan2020music,jin2024audio}. Even modest temporal offsets ($\approx$100–200ms) affect the perceived correspondence between gesture and sound. To improve synchronization and cue isolation, some recent models adopt preprocessing strategies like beat-based segmentation~\cite{10.1145/2964284.2973795} and harmonic-percussive separation~\cite{fitzgerald2010harmonic}, which can help surface rhythmically or acoustically meaningful content for downstream reasoning.

\paragraph{Tokenization strategies benefit from musical adaptation.}
Tokenization plays a central role in structuring inputs for multimodal reasoning, and recent Music AVQA models often tailor their strategies to preserve musical structure. For audio, models such as \textsc{Amuse}, \textsc{DG-SCT}, and \textsc{PSTP-Net} transform waveforms into Mel-spectrograms, which are then segmented via patch-based encoders like AST~\cite{gong2021ast} and HTS-AT or CNNs such as VGGish~\cite{45857} and ResNet-18~\cite{he2016deep}. \textsc{Audio Flamingo} \cite{10.5555/3692070.3693076}, for instance, uses overlapping 7-second windows in \textsc{ClapCap}~\cite{10448504} to embed long-range audio context. Visual streams are frequently tokenized using ViT~\cite{radford2021learning} or Swin-based~\cite{liu2022swin} patch embeddings (e.g., in \textsc{AVSiam}~\cite{lin2024siamese} and \textsc{LAVisH}~\cite{lin2023vision}), while earlier models like \textsc{AVST}~\cite{9879157} use frame-level CNN features. Text tokenization is typically handled by subword models aligned with large language models (e.g., \textsc{LLaMA2}~\cite{touvron2023llama}, \textsc{RoBERTa}~\cite{zhuang-etal-2021-robustly}), as seen in \textsc{ChatBridge} \cite{zhao2023chatbridge} and \textsc{OneLLM} \cite{han2024onellm}. These tokenization schemes help preserve temporal granularity and modality alignment, which may be important for interpreting overlapping instruments, rhythmic changes, and localized visual cues.

\paragraph{Musical content introduces distinct data and representational considerations.}
Music AVQA tasks often involve long-form performances with overlapping sources and evolving musical dynamics, which can create challenges for segmentation, annotation, and generalization. Unlike typical AVQA datasets centered on short clips and isolated actions, music-focused benchmarks (e.g., MUSIC-AVQA~\cite{9879157}) include multi-instrument performances spanning several minutes. These conditions place greater demands on dataset diversity to avoid overfitting to genre-specific patterns or ensemble configurations. To broaden coverage, some models are trained on data drawn from live performances, studio recordings, and synthetic renderings. However, the absence of symbolic structure can limit the model’s access to mid-level grounding. In this context, musically informed preprocessing (e.g., onset alignment, rhythmic segmentation, graph representation learning~\cite{huang2024exploring}) may support more interpretable and temporally aligned input representations.

\section{A Call on \textit{Specialized Spatial-Temporal Designs} for Music AVQA}
\label{sec:Architecture_influence}
We systematically analyze the models listed in Table~\ref{tab:methods} to identify architectural factors associated with strong Music AVQA performance across diverse multimodal designs. Each model is annotated based on whether it incorporates \textbf{spatial-temporal design}, defined as architectural components explicitly aimed at localizing audio-visual content in space and time—such as temporal segment selection, spatial attention, or cross-modal alignment modules. This categorization enables us to assess whether high-performing models exhibit structural traits aligned with the temporally continuous and spatially layered nature of musical performances.

\begin{figure}[ht]
  \centering
    \subfigure[\tiny Audio Counting QA.]{%
        \includegraphics[width=0.28\linewidth]{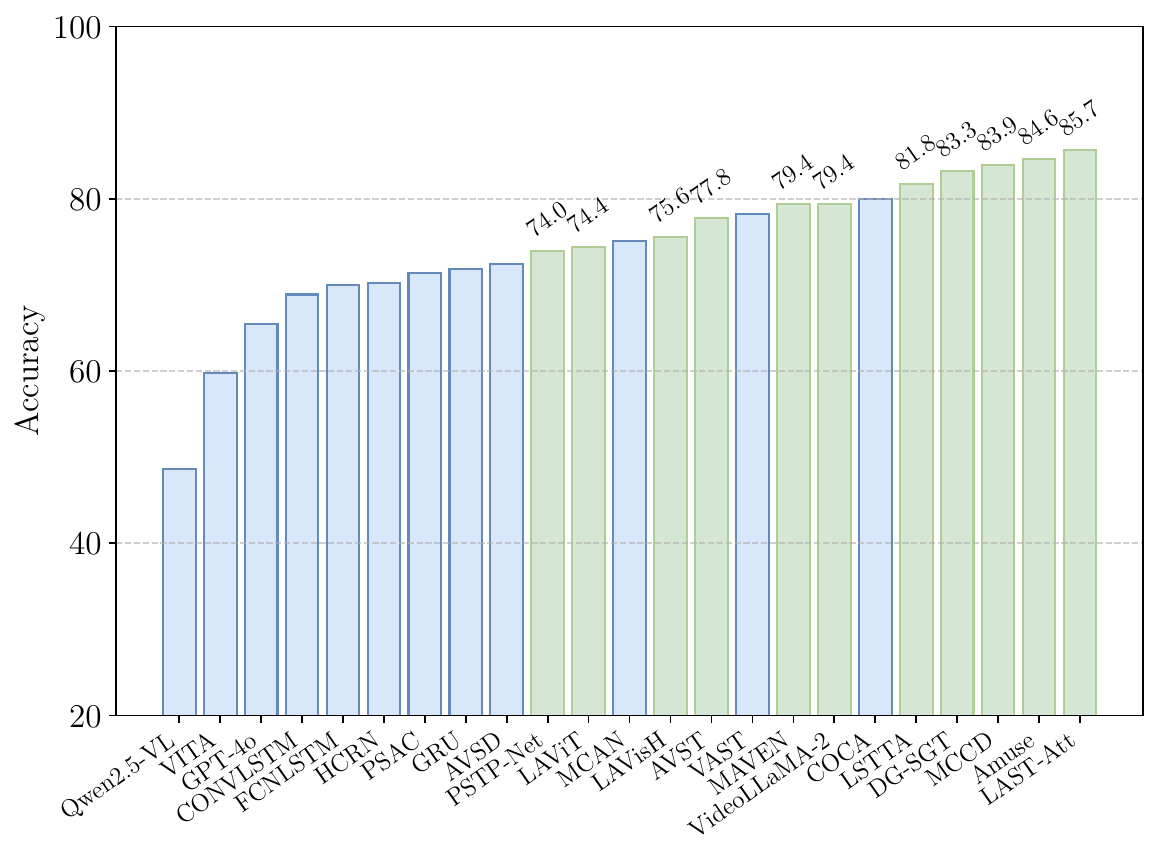}
        \label{subfig:a-counting}
    }
    \subfigure[\tiny Audio QA Average.]{%
        \includegraphics[width=0.28\linewidth]{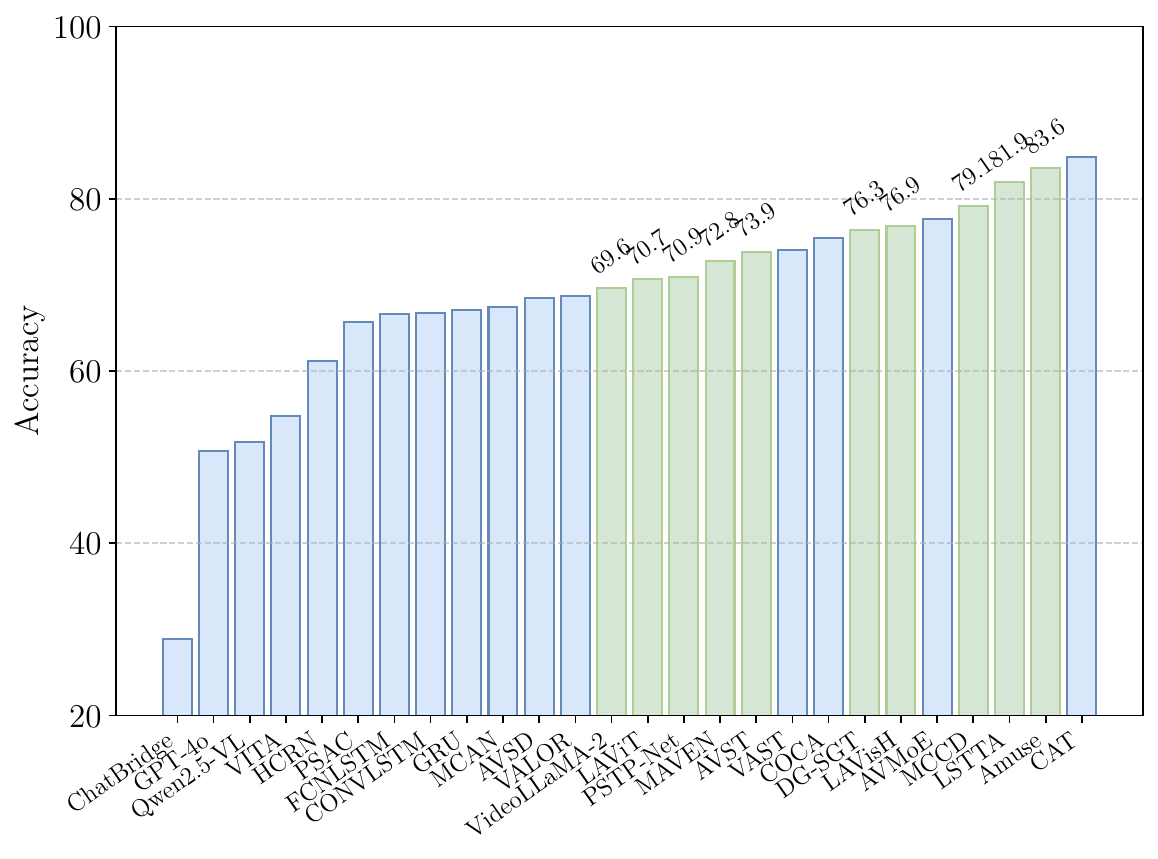}
        \label{subfig:a-average}
    }
    \subfigure[\tiny Visual Counting QA.]{%
        \includegraphics[width=0.28\linewidth]{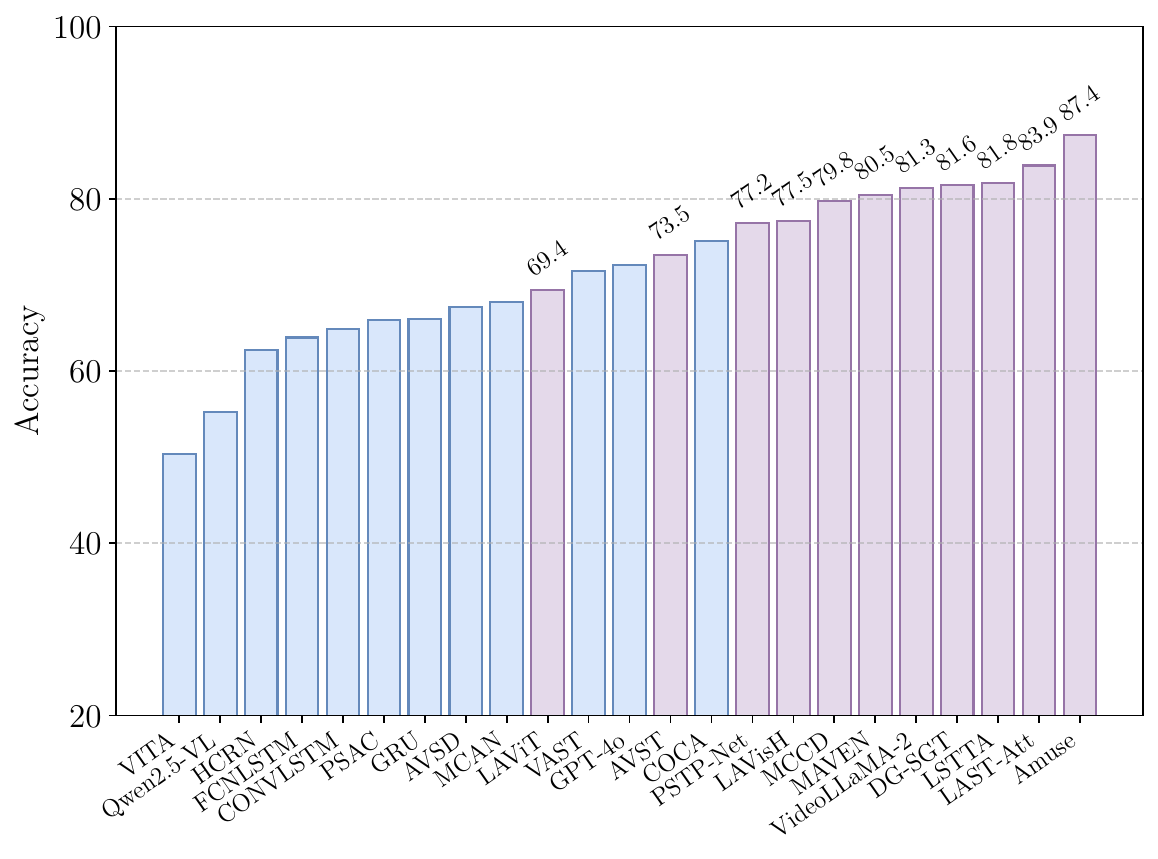}
        \label{subfig:v-counting}
    }\\[-5pt] 

    \subfigure[\tiny Visual Location QA.]{%
        \includegraphics[width=0.28\linewidth]{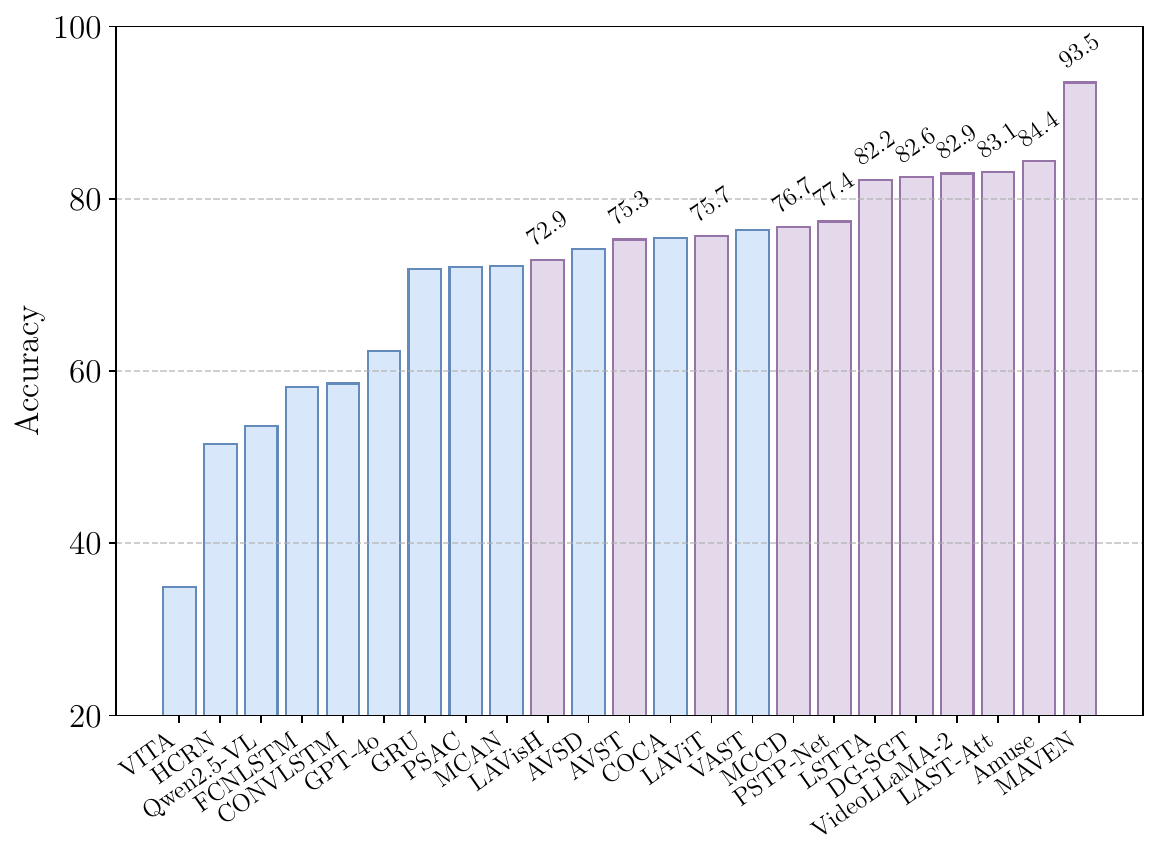}
        \label{subfig:v-location}
    }
    \subfigure[\tiny Visual QA Average.]{%
        \includegraphics[width=0.28\linewidth]{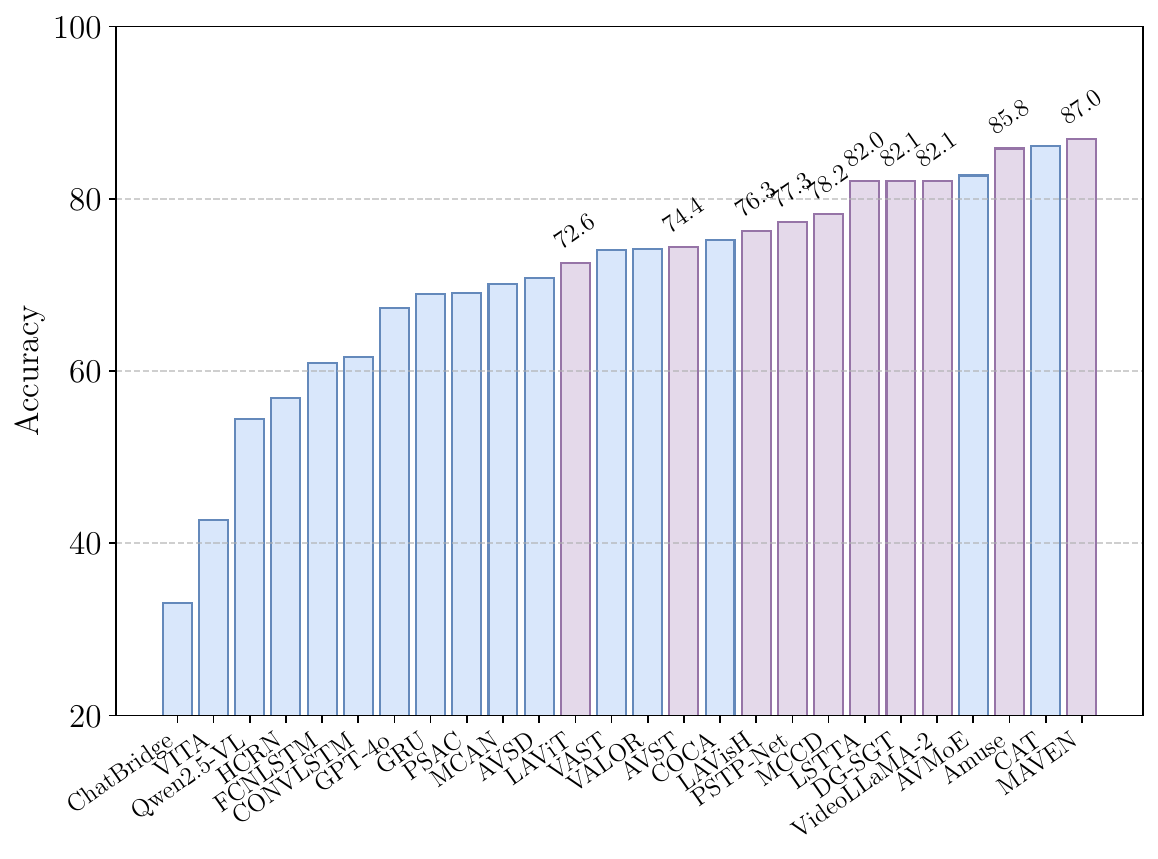}
        \label{subfig:v-average}
    }
    \subfigure[\tiny AV Location QA.]{%
        \includegraphics[width=0.28\linewidth]{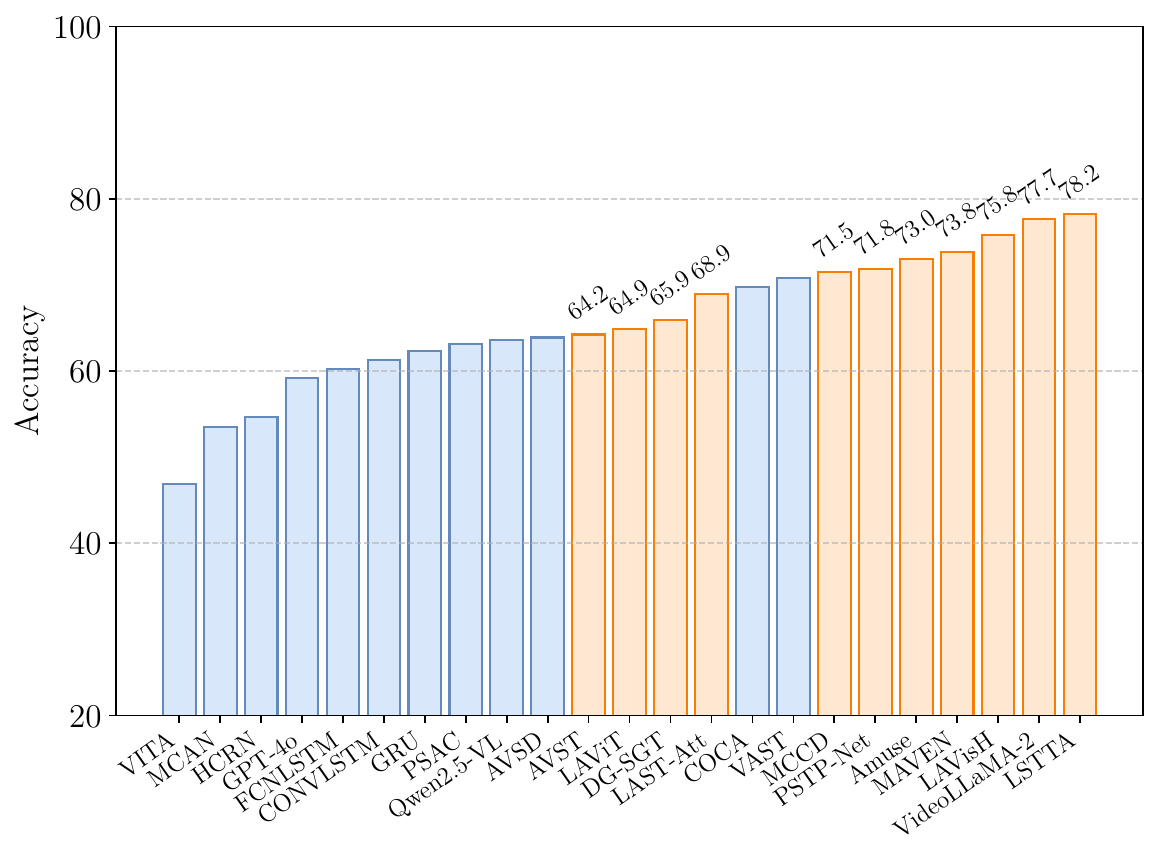}
        \label{subfig:av-location}
    }\\[-5pt]

    \subfigure[\tiny AV Comparative QA.]{%
        \includegraphics[width=0.28\linewidth]{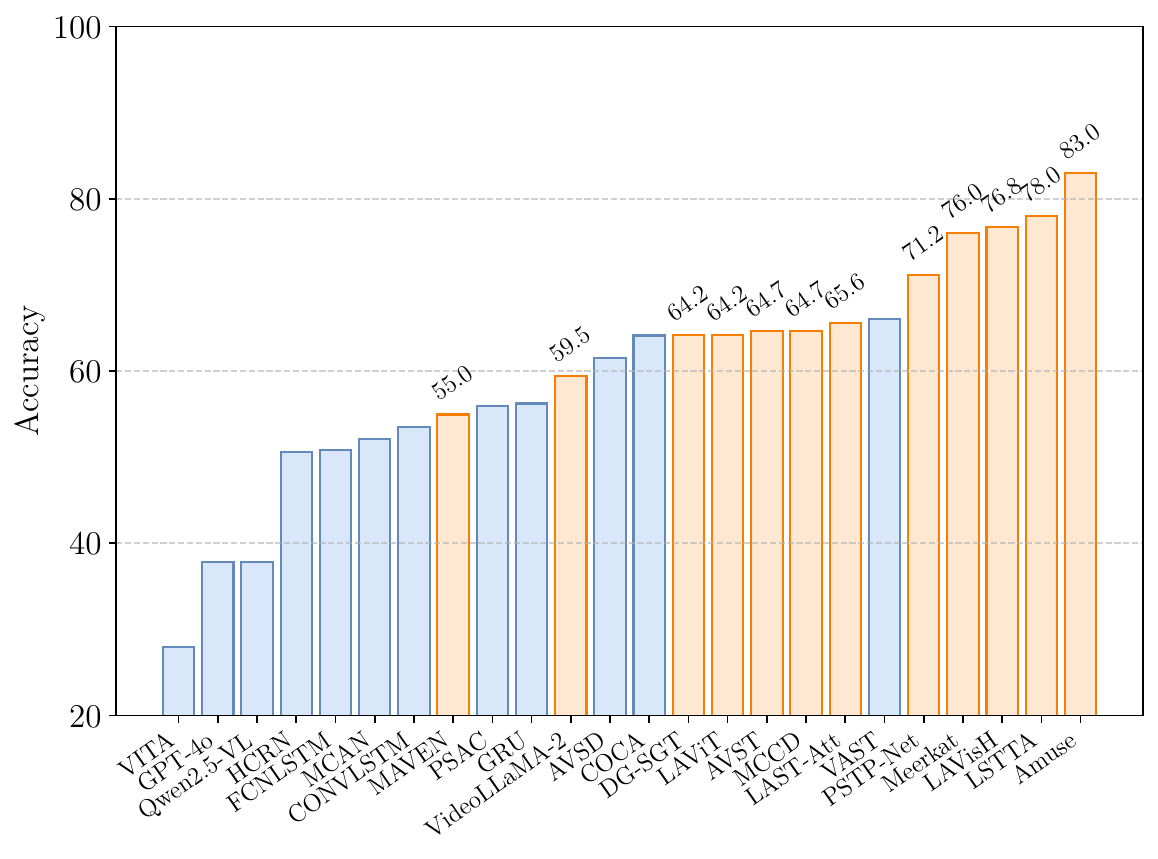}
        \label{subfig:av-comparative}
    }
    \subfigure[\tiny AV Temporal QA.]{%
        \includegraphics[width=0.28\linewidth]{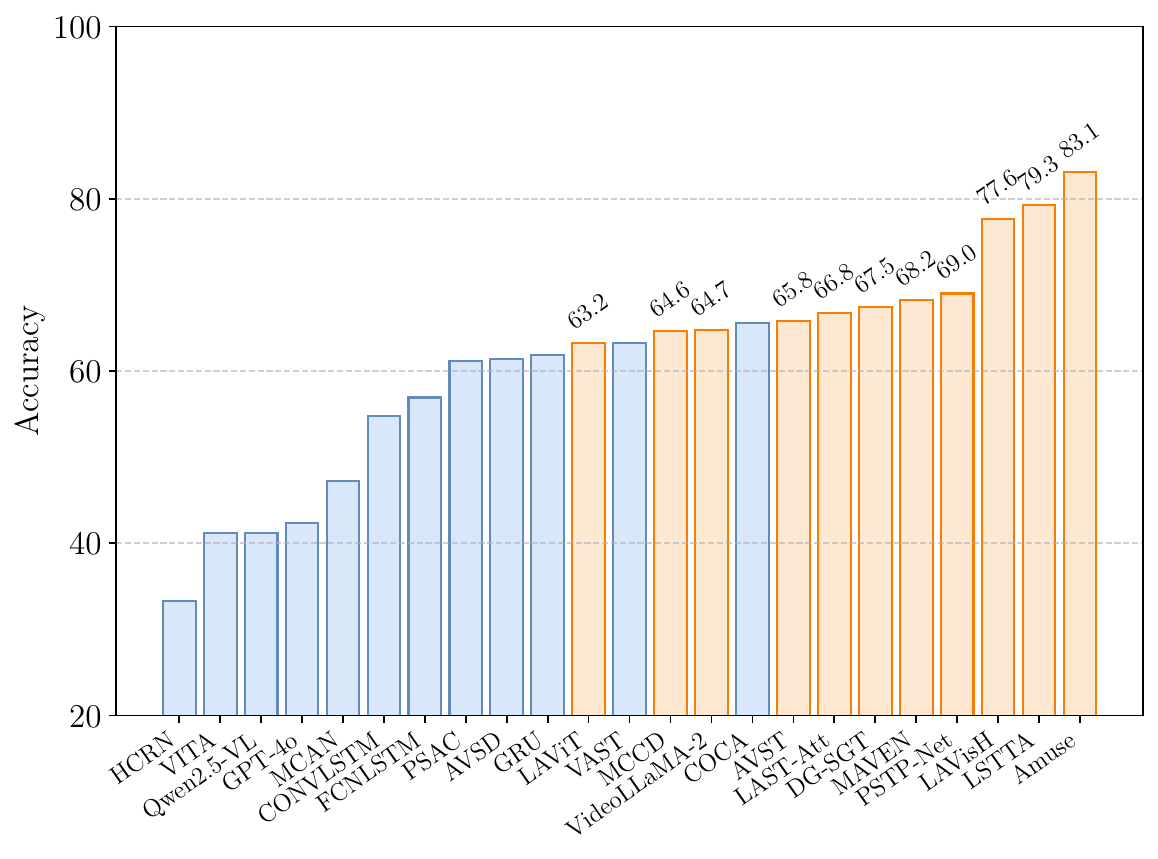}
        \label{subfig:av-temporal}
    }
    \subfigure[\tiny AV QA Average.]{%
        \includegraphics[width=0.28\linewidth]{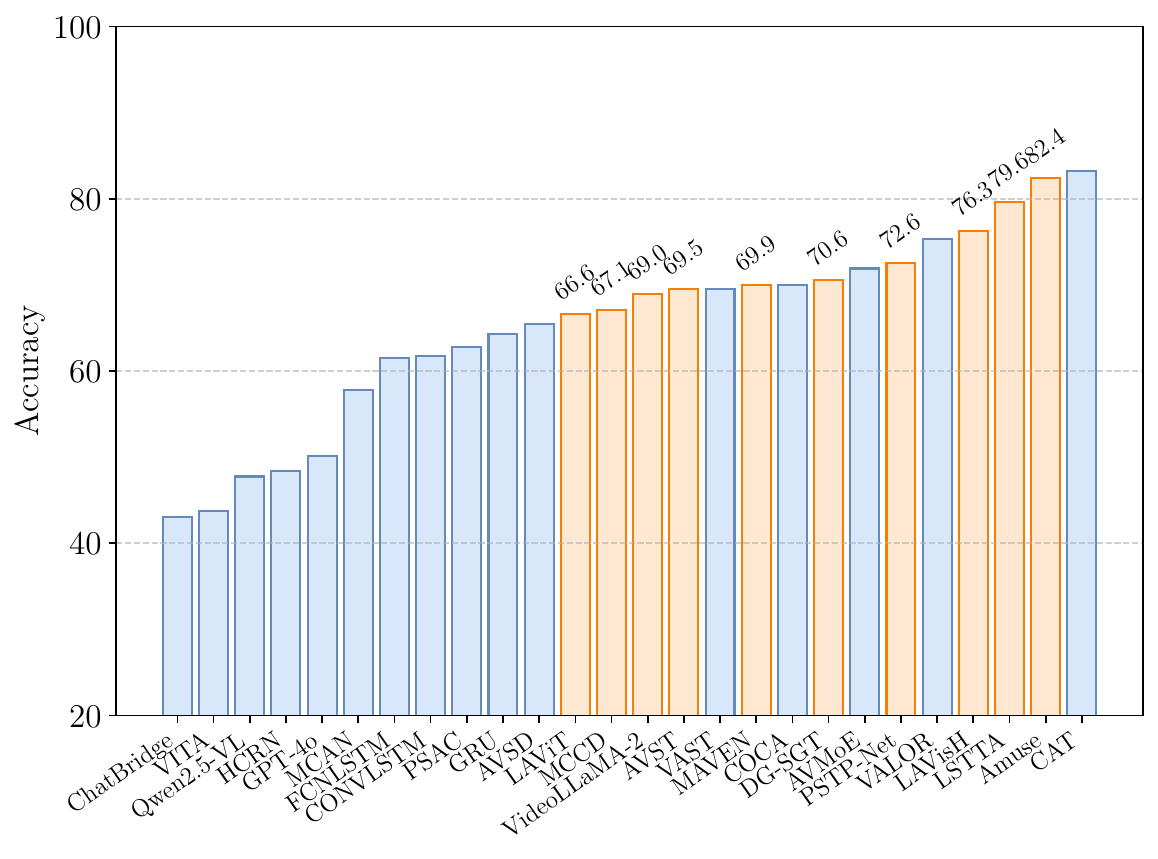}
        \label{subfig:av-average}
    }\\[-5pt]

    \vspace{-2pt}

\caption{
Accuracy comparison of Music AVQA models across representative question types, grouped by modality: (a–b) Audio, (c–e) Visual, and (f–i) Audio-Visual. Each bar corresponds to a model and is color-coded based on whether it incorporates \textbf{spatial-temporal design}: bars in ~\colorbox[HTML]{d5e7d4}{green}, ~\colorbox[HTML]{e4d9ea}{purple}, and ~\colorbox[HTML]{ffe8d2}{orange} represent models that apply spatial-temporal modeling to Audio-related, Visual-related, and Audio-Visual-related question answering, respectively; bars in ~\colorbox[HTML]{d9e7fb}{blue} represent models without spatial-temporal design. Across most categories, models with spatial-temporal components tend to perform more accurately, particularly on tasks requiring temporal reasoning or spatial localization. 
}
\label{fig:spatial-analysis}
\end{figure}

\begin{figure}[t]
    \centering
    \subfigure[\tiny Methods on Music-AVQA.]{ \includegraphics[width=0.46\linewidth]{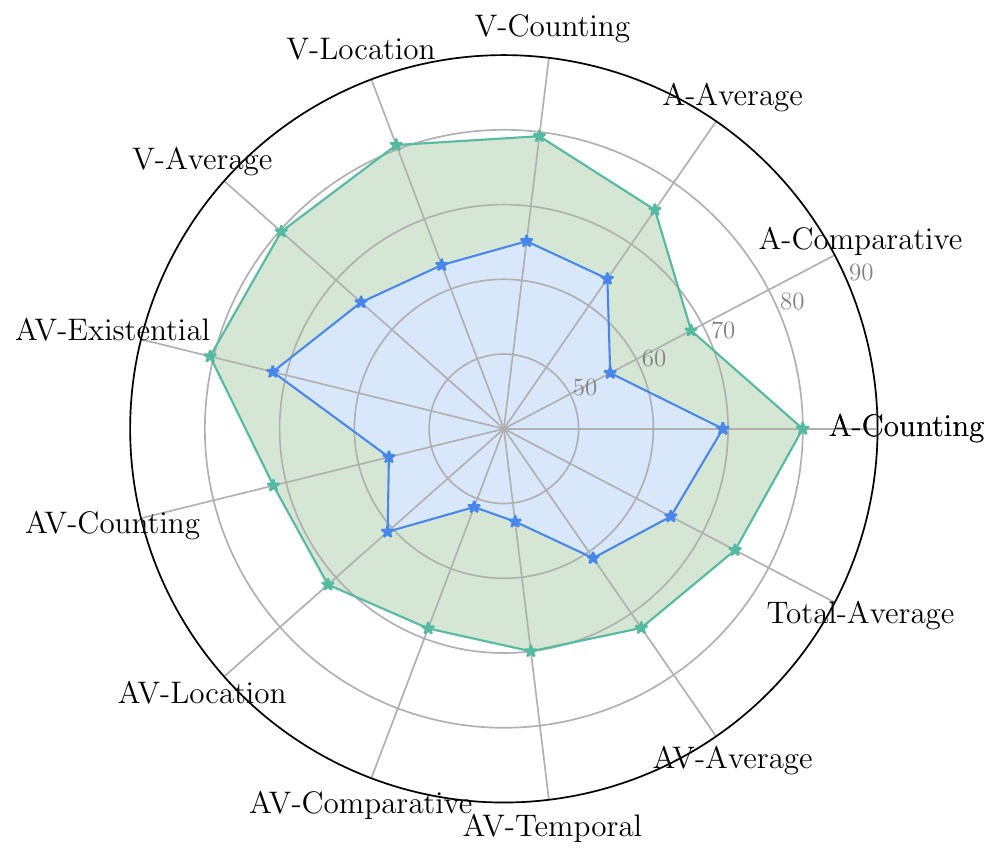}
    \label{subfig:music-avqa-radar}
    }
    \subfigure[\tiny Methods on Music-AVQA-R.]{\includegraphics[width=0.46\linewidth]{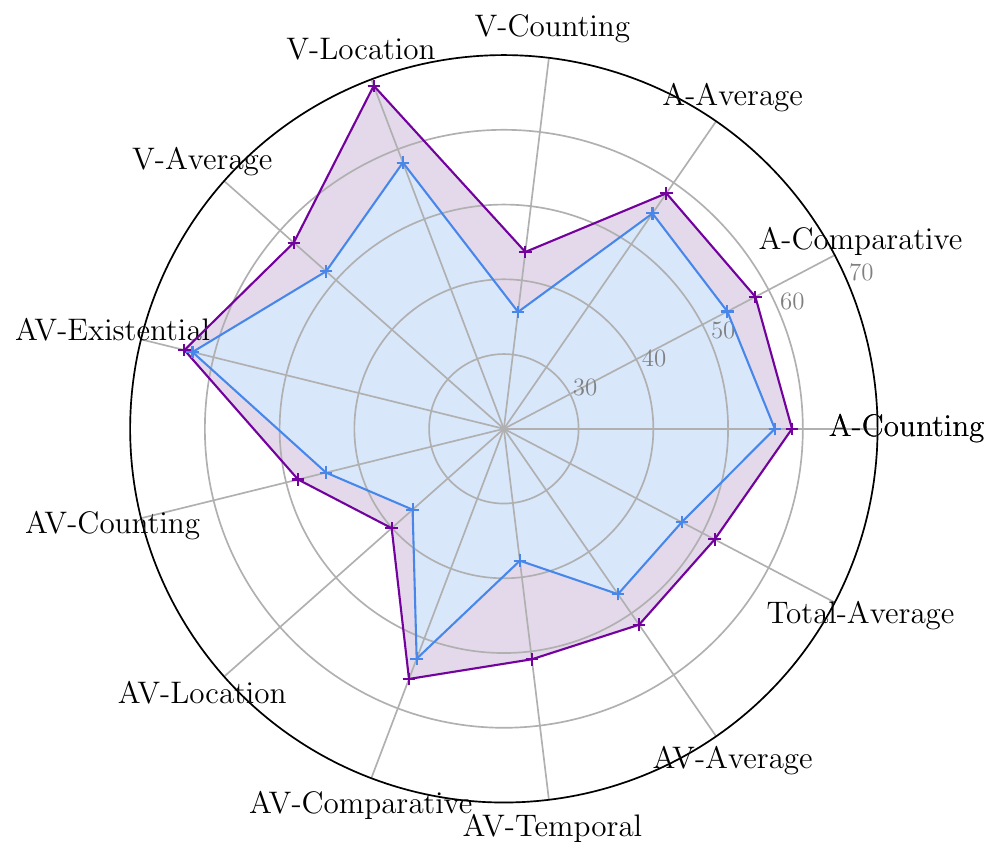}
    \label{subfig:music-avqa-r-radar}
    }
    \vspace{-0.2cm}
\caption{
Radar plots showing the per-type average accuracy of model groups with and without \textbf{spatial-temporal design} across 13 QA categories on (a) Music-AVQA~\cite{9879157} and (b) Music-AVQA-R~\cite{ma2024look}. Each axis corresponds to a QA type spanning audio, visual, and audio-visual reasoning, including the overall average (Total-Average). The filled ~\colorbox[HTML]{d5e7d4}{green} polygon in Figure \ref{subfig:music-avqa-radar} and ~\colorbox[HTML]{e4d9ea}{purple} polygon in Figure \ref{subfig:music-avqa-r-radar} represent the mean accuracy across QA types for models with spatial-temporal design, while the~\colorbox[HTML]{d9e7fb}{blue} polygon represents the average performance of models without such design. Models with spatial-temporal design consistently achieve higher accuracy. 
}
\vspace{-0.7cm}
\label{fig:radar}
\end{figure}

To assess the empirical impact of spatial-temporal design, we evaluate Music AVQA models across representative question types grouped by modality—audio, visual, and audio-visual—as shown in Figure~\ref{fig:spatial-analysis}. Each subplot compares model accuracy on a specific QA type, with bars color-coded to indicate whether spatial-temporal design is applied for the relevant modality. This setup allows precise attribution of performance differences to design choices. To capture broader trends, Figure~\ref{fig:radar} summarizes average accuracy across all 13 QA categories using radar plots on two benchmarks: Music-AVQA and Music-AVQA-R. These visualizations reveal that models with spatial-temporal design consistently outperform their counterparts, particularly in tasks involving fine-grained localization or temporal sequencing. The full quantitative results supporting these figures are reported in Appendix~\ref{appendix:quantitative-comp}, Tables~\ref{tab:acc-music-avqa}, \ref{tab:acc-music-avqa-v2}, and \ref{tab:acc-music-avqa-r}. This experimental design enables systematic assessment of spatial-temporal design as a key architectural driver of multimodal reasoning in musical environments.

\textbf{Spatial-temporal design enhances audio QA by supporting fine-grained tracking of overlapping sources and temporal acoustic variation.} Audio-related questions in Music AVQA—such as instrument counting or loudness comparison—require models to distinguish simultaneous sound sources, localize temporal onsets, and resolve temporal variation. As shown in Figures~\ref{subfig:a-counting} and~\ref{subfig:a-average}, models with spatial-temporal design consistently outperform others. \textsc{LAST-Att} achieves the highest audio counting accuracy at 85.71\%, using repeated cross-attention between question-guided Swin-Transformer features and spectrogram patches from an Audio Spectrogram Transformer to focus on musically salient moments. \textsc{Amuse}, with 83.58\% average audio QA accuracy, aligns audio-video streams using beat-synchronous features and temporally-adaptive fusion modules, allowing it to isolate relevant auditory content even under polyphonic conditions. \textsc{DG-SCT} further introduces bidirectional attention layers across temporal, spatial, and channel dimensions, dynamically adjusting audio-visual focus by question semantics. By contrast, models lacking spatial-temporal structure—such as \textsc{MCAN} (67.47\%) and \textsc{CONVLSTM} (66.73\%)—often rely on global feature pooling or frame-agnostic fusion, making them vulnerable to overlap, misalignment, and temporal drift. Notably, spatial-temporal designs adopt recurring architectural motifs: temporal segment selection (\textsc{PSTP-Net}, \textsc{AVST}), audio-guided visual attention (\textsc{DG-SCT}, \textsc{LSTTA}), and fine-grained cross-modal alignment (\textsc{Meerkat}). These mechanisms suit for modeling music’s complex structure, where overlapping instruments and evolving rhythms require localized reasoning in both time and space. The strong performance of spatial-temporal models across audio QA tasks confirms their value in resolving multi-instrument scenarios and detecting temporally grounded acoustic attributes.

\textbf{Spatial-temporal design improves visual QA by enhancing spatial disambiguation and capturing motion cues over time.} Visual-related questions in Music AVQA—such as counting instruments or identifying positions—often involve tracking multiple performers, detecting visual cues of articulation (e.g., bowing, striking), and resolving spatial relationships within densely packed frames. As shown in Figures~\ref{subfig:v-counting}–\ref{subfig:v-average}, models with spatial-temporal components generally achieve stronger accuracy. For example, \textsc{LSTTA} (82.03\% visual QA average) combines short-term semantic interaction and long-term semantic filtering modules to capture both local gestures and global scene dynamics, enabling precise reasoning about when and where instruments are engaged. \textsc{DG-SCT} (82.08\%) uses cross-modal temporal attention guided by audio prompts to enhance visual token selection, focusing on visually active regions corresponding to sounding instruments. \textsc{PSTP-Net} (77.26\%) implements a region refinement module that explicitly filters visual patches within question-relevant segments, improving spatial disambiguation. While spatial-temporal modeling is effective, some models without it still perform competitively—most notably \textsc{CAT} (86.10\%), which leverages large-scale pretrained vision encoders (ImageBind) and LLaMA2 to infer structure implicitly. However, such models may rely heavily on correlation learned from pretraining, rather than explicit reasoning about visual dynamics. Spatial-temporal models, by contrast, explicitly model the temporal unfolding of gestures and the spatial focus of performer activity—important properties in musical scenes where instrument positions are static but their activation varies over time. These architectural patterns help stabilize attention and reduce confusion when multiple instruments are visually present but only some are active, contributing to more consistent visual QA performance across counting and localization tasks.

\textbf{Spatial-temporal design is critical for audio-visual QA, where accurate reasoning requires precise temporal and spatial alignment between modalities.} Among all Music AVQA categories, audio-visual questions impose the strongest demand on cross-modal synchronization, requiring the model to associate specific acoustic events with their visual sources over time. As shown in  Figures~\ref{subfig:av-location}–\ref{subfig:av-average} and Table~\ref{tab:acc-music-avqa}, models with spatial-temporal components consistently achieve higher accuracy across AV-Existential, AV-Counting, AV-Location, AV-Comparative, and AV-Temporal types. \textsc{Amuse} reaches 82.43\% on overall AV questions by leveraging segment-level alignment between synchronized beat-level audio and video inputs and applying cross-modal adapters at each step. \textsc{PSTP-Net} adopts a progressive three-stage pipeline: temporal segment selection, spatial region refinement, and audio-guided attention, resulting in 72.57\% AV average. \textsc{Meerkat} further enhances local alignment by explicitly modeling cross-modal transport between audio patches and visual regions, and enforces bounding box constraints for grounding, yielding strong performance on AV-Comparative and AV-Location. In contrast, models without spatial-temporal design—such as \textsc{MCAN} (57.80\%), \textsc{GPT-4o} (50.08\%), and \textsc{Qwen2.5-VL} (47.75\%)—struggle to resolve fine-grained multimodal relationships. While \textsc{CAT} achieves 83.20\% AV average through large-scale pretrained encoders, its performance drops on AV-Temporal and AV-Location tasks that require precise temporal ordering or spatial binding. These results support that spatial-temporal designs—especially those involving temporally segmented reasoning, audio-guided spatial focus, and per-frame fusion—enable the model to track which instrument is sounding, when, and in which location, which is critical for answering questions such as “Did the cello on the left play after the drum on the right?”. Without such structure, models tend to conflate co-occurring signals or miss temporally offset actions, leading to lower accuracy in complex cross-modal scenarios.

\textbf{Spatial-temporal design provides a generalizable advantage across diverse Music AVQA tasks.} Our analysis reveals that models equipped with spatial-temporal design, such as beat-synchronous segment alignment in \textsc{Amuse}, progressive temporal-spatial filtering in \textsc{PSTP-Net}, and audio-guided token selection in \textsc{DG-SCT}—achieve consistently higher accuracy across audio (e.g., \textsc{LAST-Att}: 85.71\%), visual (e.g., \textsc{LSTTA}: 82.03\%), and audio-visual (e.g., \textsc{Amuse}: 82.43\%) question types. These performance gains are particularly pronounced on tasks requiring temporal ordering or cross-modal localization, as shown in Figures~\ref{fig:spatial-analysis} and~\ref{fig:radar}. Despite some strong baselines using large-scale pretrained encoders, we observe that models lacking spatial-temporal design struggle with tasks requiring temporal resolution or spatial grounding. Notably, many high-performing models adopt a common architectural pattern: (1) identifying question-relevant time segments, (2) focusing on spatial regions associated with sound cues, and (3) fusing modalities with fine-grained temporal awareness. This recurring design motif underscores spatial-temporal design as not only empirically effective, but also structurally aligned with the demands of reasoning over continuous, densely layered music data.

\section{A Call on \textit{Specialized Musical Designs} for Music AVQA}
\label{sec:musical_designs}
Current Music AVQA models typically treat musical audio as generic acoustic input, operating directly on spectrograms or waveforms without incorporating structured musical attributes such as tempo, downbeats, key, or chord progressions. More fundamentally, human understanding of music relies on hierarchical temporal structure, harmonic organization, and latent causal intent—all of which are shaped by domain-specific knowledge and perceptual priors. Inspired by this observation, we argue that musical audio should not be treated as a raw signal alone, but as a richly structured modality requiring \textit{embed musical priors and inductive structure} into models.

\paragraph{Incorporating fine-grained musical event cues.}
To support precise temporal reasoning over musical events—such as the entrance or exit of specific instruments—models can benefit from auxiliary timestamp supervision derived from musically meaningful proxies. For example, combining waveform peak analysis, Mel-frequency cepstral coefficients (MFCCs), and spectral change detection can help identify dynamic shifts in the audio stream. Beat-tracking algorithms (e.g., from Librosa) can segment audio by rhythm, while pitch-based estimators (e.g., Aubio's YIN) can trace changes in dominant frequency to indicate evolving instrumental activity. These mid-level cues can be used to generate pseudo-labels for training timestamp encoders, enabling models to better localize temporally anchored events. Embedding such representations into Music AVQA pipelines may improve event-level understanding and enhance the interpretability of the model’s temporal predictions.

\paragraph{Embedding mid-level musical structure into multimodal models.}
Structured musical features (tempo, key, downbeats, chord progressions) provide a coherent framework for aligning audio-visual inputs across time. Symbolic or MIR-derived signals yield interpretable, temporally smooth trajectories that reflect music's hierarchical organization (phrases, sections, transitions). They abstract from low-level waveform fluctuations and provide a musically meaningful scaffold that persists across genres, tempos, and instrumentation. Integrating them as auxiliary inputs or attention-guiding signals can improve capture of long-range dependencies, maintenance of rhythmic continuity, and resolution of ambiguous instrument interactions, especially in polyphonic or ensemble contexts. This structured conditioning serves as a musical inductive bias, particularly helpful in complex multimodal scenes where overlapping sources challenge bottom-up fusion strategies and where salient events may not be visually or acoustically distinct without temporal alignment cues.

\paragraph{Modeling latent musical reasoning trajectories.}
Many Music AVQA questions hinge on implicit causal or temporal relations (e.g., who initiated a phrase, whether an entrance shifted ensemble balance). They often lack step-level supervision, so labels alone do not reveal reasoning paths. We introduce latent reasoning trajectories: structured internal variables representing evolving hypotheses about the musical scene. Rather than mapping inputs to answers, the model infers intermediate latent states (e.g., current lead instrument, rhythmic-intensity change, impending entrances) and updates them as multimodal evidence arrives. Architecturally, this can be realized with hierarchical latent-variable models or recurrent variational modules, with states encoding musical intentions, transitions, and causal flow. These hidden trajectories simulate plausible event sequences, enabling extrapolation and bridging missing links between observed signals. This latent reasoning improves generalization by embedding inductive structure aligned with human inference of musical cause and progression, beyond surface audio-visual co-occurrence.

\paragraph{Supervising chain-of-thought reasoning in musical QA.}
Some musical questions, especially with temporal or causal dependencies, require sequential sub-decisions to reach the correct answer. For example, ``Which instrument enters after the piano stops?'' involves (1) detecting the piano stop, (2) detecting subsequent onsets, and (3) selecting the earliest new instrument. Rather than black-box classification, models can be trained to emit intermediate steps via supervised rationales or pseudo-labels from MIR event detection. This approach, akin to chain-of-thought (CoT) in LLMs, improves transparency, supports modular subgoals, and maintains cross-modal alignment. Stepwise supervision highlights failure modes in temporal or semantic inference, providing clearer diagnostics. In music context, CoT chains can include domain-specific steps such as beat alignment, timbre matching, or onset attribution. These interpretable traces support higher accuracy for multi-stage queries and expose reasoning shortcuts and dataset biases.

\section{Conclusion}
\label{sec:conclusion}
This paper underscores that successful Music AVQA necessitates specialized designs tailored to the unique demands of musical content: fine-grained audio-visual processing, precise spatial-temporal modeling, and integrated domain knowledge. We call on the research community to: (i) develop more nuanced music understanding benchmarks, and (ii) explore hybrid architectures combining MLLM strengths with music-specific components. The future of effective multimodal AI lies not in a universal approach, but in the thoughtful integration of general capabilities with deep, domain-specific expertise, benefiting music understanding and other complex fields.

\section*{Limitations}
This paper has several limitations. First, as Music AVQA continues to evolve rapidly, some very recent works may not be fully reflected in this manuscript. Second, although we aim to provide a clear and structured synthesis of existing datasets, methods, and future directions, our organization of the field reflects one possible perspective, and alternative categorizations may also be reasonable. Third, our discussion mainly emphasizes model designs and reasoning patterns, and therefore does not cover aspects such as data collection and human evaluation in equal depth.

\section*{Ethical Considerations}
All experiments presented in this study were conducted using publicly available datasets and models licensed for academic research purposes. To the best of our knowledge, this work does not present any ethical concerns.

\bibliography{ref}

\clearpage           
\appendix

\startcontents[appendices]
\section*{Contents of Appendix}
\printcontents[appendices]{}{1}{\normalsize}

\section{Quantitative Comparison on Music AVQA Datasets}\label{appendix:quantitative-comp}
We present comprehensive quantitative comparisons of recent state-of-the-art methods on multiple Music AVQA datasets~\cite{9879157,10484352,ma2024look}, shown in Table~\ref{tab:acc-music-avqa},~\ref{tab:acc-music-avqa-v2}, and~\ref{tab:acc-music-avqa-r}. 
We evaluate the models across a diverse set of question categories, spanning Audio-related, Visual-related, and Audio\&Visual-related reasoning tasks.
For each dataset, we report accuracy metrics for subcategories such as Counting, Comparative, Location, Existential, and Temporal reasoning, 
along with average accuracy within each modality and the overall performance.

\begin{table*}[htbp]
    \centering
\small
\resizebox{1.0\textwidth}{!}{
    \begin{tabular}{p{3.8cm}p{0.9cm}p{0.9cm}p{0.9cm}p{0.9cm}p{0.9cm}p{0.9cm}p{0.9cm}p{0.9cm}p{0.9cm}p{0.9cm}p{0.9cm}p{0.9cm}p{0.9cm}}
        \toprule
        \multirow{2}{*}{\textbf{Methods}}
        & \multicolumn{3}{c}{\textbf{Audio-related QA}}
        & \multicolumn{3}{c}{\textbf{Visual-related QA}}
        & \multicolumn{6}{c}{\textbf{Audio\&Visual-related QA}}
        & \multirow{2}{*}{\textbf{Avg}} \\   
        \cmidrule(lr){2-4}
        \cmidrule(lr){5-7}
        \cmidrule(lr){8-13}
        & \textbf{Count} & \textbf{Comp} & \textbf{Avg}
        & \textbf{Count} & \textbf{Local} & \textbf{Avg}
        & \textbf{Exist} & \textbf{Count} & \textbf{Local} & \textbf{Comp} & \textbf{Temp}
        & \textbf{Avg}\\                  
        \midrule
    \rowcolor{gray!15}\textsc{Amuse}~\cite{diao2024learning}&	84.61&	82.45&	83.58&	87.41&	84.39&	85.84&	86.95&	85.49&	73.01&	82.98&	83.06&	82.43&	83.52\\
    \textsc{Audio Flamingo}~\cite{10.5555/3692070.3693076}&	-&	-&	-&	-&	-&	-&	-&	-&	-&	-&	-&	-&	-\\
\textsc{AVMoE}~\cite{cheng2025mixtures}&	-&	-&	77.60&	-&	-&	82.70&	-&	-&	-&	-&	-&	71.90&	75.70\\
\textsc{AVSD}~\cite{schwartz2019simple}&	72.41&	61.90&	68.52&	67.39&	74.19&	70.83&	81.61&	58.79&	63.89&	61.52&	61.41&	65.49&	67.44\\
\textsc{AVSiam}~\cite{lin2024siamese}&	-&	-&	-&	-&	-&	-&	-&	-&	-&	-&	-&	-&	-\\
\rowcolor{gray!15}\textsc{AVST}~\cite{9879157}&	77.78&	67.17&	73.87&	73.52&	75.27&	74.40&	82.49&	69.88&	64.24&	64.67&	65.82&	69.53&	71.59\\
%%%%%%%%%%%%%%%%%%%%%%%%
\textsc{CAT}~\cite{ye2024cat}&	-&	-&	84.90&	-&	-&	86.10&	-&	-&	-&	-&	-&	83.20&	84.30\\
\textsc{ChatBridge}~\cite{zhao2023chatbridge}&	-&	-&	28.90&	-&	-&	33.10&	-&	-&	-&	-&	-&	43.00&	78.90\\
\rowcolor{gray!15}\textsc{CIGN}~\cite{mo2023class}&	-&	-&	-&	-&	-&	-&	-&	-&	-&	-&	-&	-&	-\\
\textsc{COCA}~\cite{lao2023coca}&	79.94&	67.68&	75.42&	75.10&	75.43&	75.23&	83.50&	66.63&	69.72&	64.12&	65.57&	69.96&	72.33\\
\textsc{CONVLSTM}~\cite{9145807}&	68.88&	63.06&	66.73&	64.89&	58.55&	61.68&	82.81&	55.99&	61.30&	53.45&	54.73&	61.75&	62.61\\
\textsc{CrossMAE}~\cite{guo2024crossmae}&	-&	-&	-&	-&	-&	-&	-&	-&	-&	-&	-&	-&	-\\
%%%%%%%%%%%%%%%%%%%%%%%%%%%%
\rowcolor{gray!15}\textsc{DCL}~\cite{lv2023disentangled}&	-&	-&	-&	-&	-&	-&	-&	-&	-&	-&	-&	-&	-\\
\rowcolor{gray!15}\textsc{DG-SCT}~\cite{duan2023cross}&	83.27&	64.56&	76.34&	81.57&	82.57&	82.08&	81.61&	72.84&	65.91&	64.22&	67.48&	70.56&	74.62\\
\rowcolor{gray!15}\textsc{EEMC}~\cite{10.1007/978-3-031-72904-1_12}&	-&	-&	-&	-&	-&	-&	-&	-&	-&	-&	-&	-&	-\\
\textsc{FCNLSTM}~\cite{9145807}&	69.96&	61.06&	66.67&	63.89&	58.14&	60.98&	83.42&	56.31&	60.28&	50.85&	56.92&	61.46&	62.25\\
\textsc{GPT-4o}~\cite{hurst2024gpt}&	65.42&	36.07&	50.75&	72.36&	62.30&	67.33&	56.12&	54.84&	59.23&	37.84&	42.35&	50.08&	54.06\\
\textsc{GRU}~\cite{7410636}&	71.82&	58.90&	67.04&	66.06&	71.82&	68.97&	81.41&	60.30&	62.32&	56.23&	61.89&	64.26&	66.00\\
\textsc{HCRN}~\cite{le2020hierarchical}&70.21&	45.62&	61.14&	62.41&	51.51&	56.90&	52.94&	42.07&	54.70&	50.59&	33.33&	48.41&	52.54\\
\rowcolor{gray!15}\textsc{LAST-Att}~\cite{10484352}&	85.71&	63.10&	-&	83.86&	83.09&	-&	76.47&	76.20&	68.91&	65.60&	66.75&	-&	75.45\\
\rowcolor{gray!15}\textsc{LAVisH}~\cite{lin2023vision}&	75.59&	84.13&	76.86&	77.45&	72.91&	76.29&	71.91&	77.52&	75.81&	76.75&	77.62&	76.31&	76.10\\
\rowcolor{gray!15}\textsc{LAViT}~\cite{yun2021pano}&	74.36&	64.56&	70.73&	69.39&	75.65&	72.56&	81.21&	59.33&	64.91&	64.22&	63.23&	66.64&	68.93\\
\rowcolor{gray!15}\textsc{LSTTA}~\cite{liu2023parameter}&	81.75&	82.04&	81.90&	81.82&	82.23&	82.03&	83.46&	79.11&	78.23&	78.02&	79.32&	79.63&	81.19\\
\rowcolor{gray!15}\textsc{MAVEN}~\cite{ma2025fortisavqa}&	79.44&	54.10&	72.79&	80.49&	93.50&	86.99&	87.00&	66.67&	73.85&	54.95&	68.24&	69.94&	74.60\\
\textsc{MCAN}~\cite{8953581}&	75.05&	54.58&	67.47&	68.06&	72.15&	70.13&	81.91&	54.15&	53.45&	52.11&	47.21&	57.80&	62.77\\
\rowcolor{gray!15}\textsc{MCCD}~\cite{ma2024look}&	83.87&	71.04&	79.14&	79.78&	76.73&	78.24&	80.87&	51.63&	71.46&	64.67&	64.60&	67.13&	72.20\\
\rowcolor{gray!15}\textsc{Meerkat}~\cite{10.1007/978-3-031-73039-9_4}&	-&	-&	-&	-&	-&	-&	-&	85.70&	-&	75.98&	-&	-&	-\\
\textsc{OneLLM}~\cite{han2024onellm}&	-&	-&	-&	-&	-&	-&	-&	-&	-&	-&	-&	-&	47.60\\
\textsc{OPM}~\cite{wei2024fly}&	-&	-&	-&	-&	-&	-&	-&	-&	-&	-&	-&	-& 70.80\\
\textsc{PSAC}~\cite{10.1609/aaai.v33i01.33018658}&	71.33&	56.07&	65.68&	65.89&	72.07&	69.02&	78.59&	54.80&	63.11&	55.96&	61.17&	62.75&	64.92\\
\rowcolor{gray!15}\textsc{PSTP-Net}~\cite{li2023progressive}&	73.97&	65.59&	70.90&	77.15&	77.36&	77.26&	76.18&	73.23&	71.80&	71.19&	69.00&	72.57&	73.52\\
\textsc{QaP}~\cite{liang2024querying}&	-&	-&	-&	-&	-&	-&	-&	-&	-&	-&	-&	-& -\\
\textsc{Qwen2.5-VL}~\cite{bai2025qwen2}&	48.60&	55.00&	51.80&	55.28&	53.66&	54.47&	44.00&	52.17&	63.57&	37.84&	41.18&	47.75&	50.14\\
%%%
\rowcolor{gray!15}\textsc{RefAtomNet}~\cite{peng2024referring}&	-&	-&	-&	-&	-&	-&	-&	-&	-&	-&	-&	-&	-\\
\textsc{VALOR}~\cite{10.1109/TPAMI.2024.3479776}&	-&	-&	68.70&	-&	-&	74.20&	-&	-&	-&	-&	-&	75.30&	78.90\\
\textsc{VAST}~\cite{10.5555/3666122.3669307}&	78.18&	67.05&	74.06&	71.56&	76.38&	74.00&	81.81&	64.51&	70.80&	66.01&	63.23&	69.54&	71.52\\
\rowcolor{gray!15}\textsc{VideoLLaMA-2}~\cite{cheng2024videollama}&	79.44&	52.46&	69.64&	81.30&	82.93&	82.11&	77.00&	63.44&	77.69&	59.46&	64.71&	68.98&	72.56\\
\textsc{VITA}~\cite{fu2024vita}&	59.81&	45.90&	54.76&	50.41&	34.96&	42.68&	54.00&	49.46&	46.92&	27.93&	41.18&	43.74&	45.44\\
        \bottomrule
    \end{tabular}
    }
     \caption{Comparison with state-of-the-art methods on the Music AVQA~\cite{9879157} test set. We report the accuracy for Audio (Counting, Comparative), Visual (Counting, Location), and Audio-Visual (Existential, Counting, Location, Comparative, Temporal) question types, along with the average accuracy for Audio, Visual, Audio-Visual, and overall. Methods highlighted with a \colorbox{gray!15}{gray} background incorporate spatial-temporal designs.}
\label{tab:acc-music-avqa}
\end{table*}

\begin{table*}[htbp]
    \centering
\small
\resizebox{1.0\textwidth}{!}{
    \begin{tabular}{p{3.8cm}p{0.9cm}p{0.9cm}p{0.9cm}p{0.9cm}p{0.9cm}p{0.9cm}p{0.9cm}p{0.9cm}p{0.9cm}p{0.9cm}p{0.9cm}p{0.9cm}p{0.9cm}}
        \toprule
        \multirow{2}{*}{\textbf{Methods}}
        & \multicolumn{3}{c}{\textbf{Audio-related QA}}
        & \multicolumn{3}{c}{\textbf{Visual-related QA}}
        & \multicolumn{6}{c}{\textbf{Audio\&Visual-related QA}}
        & \multirow{2}{*}{\textbf{Avg}} \\   
        \cmidrule(lr){2-4}
        \cmidrule(lr){5-7}
        \cmidrule(lr){8-13}
        & \textbf{Count} & \textbf{Comp} & \textbf{Avg}
        & \textbf{Count} & \textbf{Local} & \textbf{Avg}
        & \textbf{Exist} & \textbf{Count} & \textbf{Local} & \textbf{Comp} & \textbf{Temp}
        & \textbf{Avg}\\                          
        \midrule
\rowcolor{gray!15}\textsc{Amuse}~\cite{diao2024learning}&	84.76&	83.88&	84.34&	88.15&	85.16&	86.74&	88.30&	87.47&	78.77&	84.41&	85.38&	85.51&	85.16\\
        \rowcolor{gray!15}\textsc{AVST}~\cite{9879157}&	81.74&	62.11&	72.46&	79.08&	77.64&	78.40&	72.12&	69.03&	65.05&	63.98&	60.57&	66.26&	71.08\\
\rowcolor{gray!15}\textsc{DG-SCT}~\cite{duan2023cross}&	83.66&	62.47&	73.64&	82.05&	82.97&	82.48&	83.43&	72.70&	64.65&	64.78&	67.34&	70.38&	74.08\\
        \rowcolor{gray!15}\textsc{LAST-Att}~\cite{10484352}&	86.03&	62.52&	-&	84.12&	84.01&	-&	76.21&	75.23&	68.91&	65.60&	60.60&	-&	75.44\\
\rowcolor{gray!15}\textsc{LAVisH}~\cite{lin2023vision}&	84.36&	58.57&	72.17&	83.25&	81.46&	82.40&	73.26&	73.45&	65.64&	64.26&	60.82&	67.75&	72.34\\
        \bottomrule
\end{tabular}
}
\caption{Comparison with state-of-the-art methods on the Music AVQA v2.0~\cite{10484352} test set. We report the accuracy for Audio (Counting, Comparative), Visual (Counting, Location), and Audio-Visual (Existential, Counting, Location, Comparative, Temporal) question types, along with the average accuracy for Audio, Visual, Audio-Visual, and overall. Methods highlighted with a \colorbox{gray!15}{gray} background incorporate spatial-temporal designs.}
\label{tab:acc-music-avqa-v2}
\end{table*}

\begin{table*}[htbp]
    \centering
\small
\resizebox{1.0\textwidth}{!}{
    \begin{tabular}{p{3.8cm}p{0.9cm}p{0.9cm}p{0.9cm}p{0.9cm}p{0.9cm}p{0.9cm}p{0.9cm}p{0.9cm}p{0.9cm}p{0.9cm}p{0.9cm}p{0.9cm}p{0.9cm}}
        \toprule
        \multirow{2}{*}{\textbf{Methods}}
        & \multicolumn{3}{c}{\textbf{Audio-related QA}}
        & \multicolumn{3}{c}{\textbf{Visual-related QA}}
        & \multicolumn{6}{c}{\textbf{Audio\&Visual-related QA}}
        & \multirow{2}{*}{\textbf{Avg}} \\   
        \cmidrule(lr){2-4}
        \cmidrule(lr){5-7}
        \cmidrule(lr){8-13}
        & \textbf{Count} & \textbf{Comp} & \textbf{Avg}
        & \textbf{Count} & \textbf{Local} & \textbf{Avg}
        & \textbf{Exist} & \textbf{Count} & \textbf{Local} & \textbf{Comp} & \textbf{Temp}
        & \textbf{Avg}\\                          
        \midrule
\textsc{Att-BLSTM}~\cite{zhou-etal-2016-attention}&	60.00&	49.55&	54.77&	32.15&	47.97&	48.89&	54.33&	39.46&	32.52&	51.00&	24.45&	40.35&	40.35\\
\textsc{AVSD}~\cite{schwartz2019simple}&50.92&	54.20&	52.56&	35.21&	68.11&	52.20&	64.13&	36.68&	27.14&	58.99&	40.83&	45.55&	45.55\\
\textsc{CONVLSTM}~\cite{9145807}&	55.68&	60.22&	57.95&	35.64&	51.66&	52.23&	72.45&	53.18&	32.35&	57.91&	43.33&	51.84&	51.84\\
\textsc{FCNLSTM}~\cite{9145807}&	51.36&	57.96&	54.66&	33.53&	52.96&	50.09&	71.64&	51.98&	34.96&	57.40&	33.90&	49.98&	49.98\\
\textsc{GRU}~\cite{7410636}&	57.78&	58.95&	58.36&	38.08&	57.67&	54.17&	70.53&	43.33&	39.70&	57.29&	35.85&	49.34&	49.34\\
\textsc{HCAttn}~\cite{10.5555/3157096.3157129}&51.65&	53.12&	52.38&	32.86&	60.09&	50.02&	63.85&	39.77&	36.01&	54.47&	36.54&	46.13&	46.13\\
\textsc{HCRN}~\cite{le2020hierarchical}&54.42&	39.81&	47.11&	32.71&	45.34&	43.88&	53.63&	39.67&	37.08&	35.10&	42.30&	41.56&	41.56\\
\textsc{HME}~\cite{fan2019heterogeneous}&	58.28&	56.63&	57.45&	33.71&	65.93&	54.40&	66.12&	39.91&	40.18&	56.89&	37.76&	48.17&	48.17\\
\rowcolor{gray!15}\textsc{LAVisH}~\cite{lin2023vision}&	52.86&	62.72&	57.79&	38.33&	67.47&	55.83&	78.65&	41.48&	32.38&	62.18&	44.05&	51.75&	51.75\\
\rowcolor{gray!15}\textsc{LAViT}~\cite{yun2021pano}&	47.01&	47.86&	47.43&	31.39&	66.35&	48.01&	37.21&	53.02&	36.87&	43.05&	42.17&	42.46&	42.46\\
\textsc{MCAN}~\cite{8953581}&	67.59&	54.49&	61.04&	45.64&	64.37&	58.62&	59.29&	53.86&	45.02&	51.49&	46.35&	51.20&	51.20\\
\rowcolor{gray!15}\textsc{MCCD}~\cite{ma2024look}&	75.78&	63.43&	69.60&	61.76&	73.43&	68.80&	76.18&	50.55&	50.92&	62.15&	66.95&	61.35&	61.35\\
\textsc{PSAC}~\cite{10.1609/aaai.v33i01.33018658}&	54.85&	52.77&	53.81&	37.99&	66.83&	53.25&	53.05&	47.14&	38.14&	48.53&	36.46&	44.66&	44.66\\
        \bottomrule
\end{tabular}
}
\caption{Comparison with state-of-the-art methods on the Music-AVQA-R~\cite{ma2024look} test set. We report the accuracy for Audio (Counting, Comparative), Visual (Counting, Location), and Audio-Visual (Existential, Counting, Location, Comparative, Temporal) question types, along with the average accuracy for Audio, Visual, Audio-Visual, and overall. Methods highlighted with a \colorbox{gray!15}{gray} background incorporate spatial-temporal designs.}
\label{tab:acc-music-avqa-r}
\end{table*}

\section{Representative Examples of Music Performance Scene Types}
\label{appendix:example-scene-types} 

\begin{figure}[ht]
    \centering
    \subfigure[Solo Performance.]{
        \includegraphics[width=0.98\linewidth]{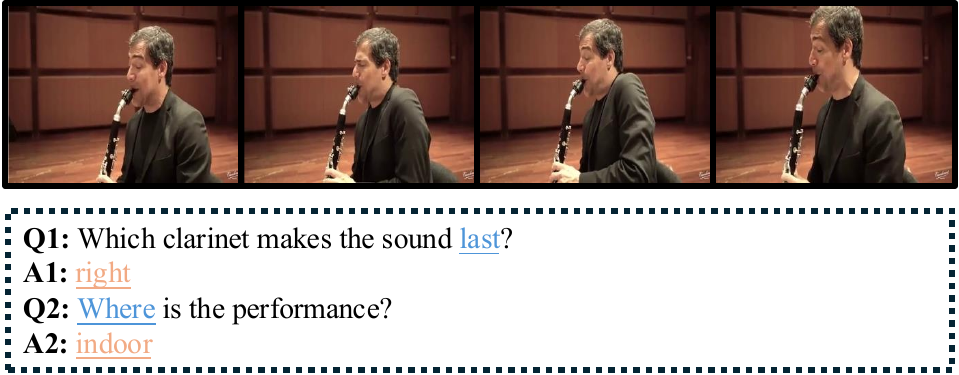}
        \label{subfig:solo-performance}
    }\\
    \subfigure[Ensemble of the Same Instrument.]{
        \includegraphics[width=0.98\linewidth]{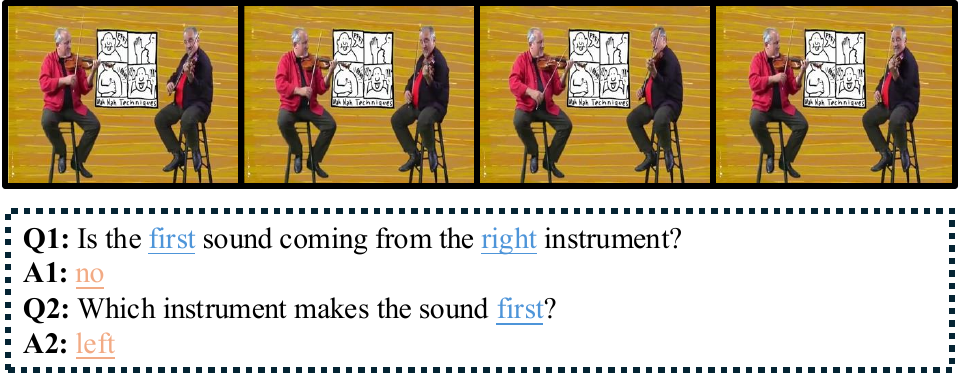}
        \label{subfig:ensemble-same}
    }\\
    \subfigure[Ensemble of Different Instruments.]{
        \includegraphics[width=0.98\linewidth]{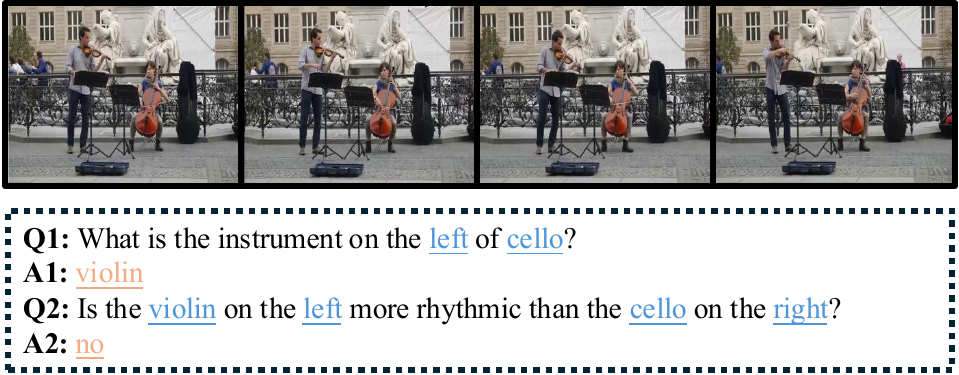}
        \label{subfig:ensemble-different}
    }\\
    \subfigure[Culture-Specific Ensemble.]{
        \includegraphics[width=0.98\linewidth]{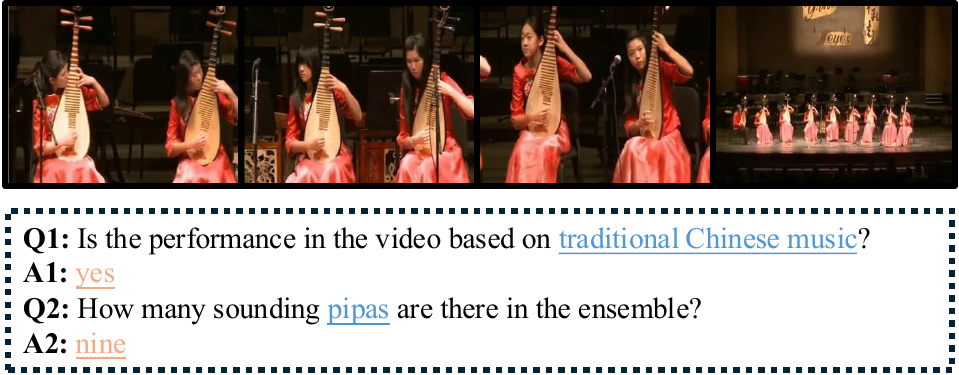}
        \label{subfig:culture-spec}
    }
\caption{
Representative examples for the four common music performance scene types.  
    (a) \emph{Solo performance}: a single musician highlights individual virtuosity on one instrument.  
    (b) \emph{Ensemble of the same instrument}: multiple players of identical (or closely related) instruments create timbral thickness and homogeneous harmony.  
    (c) \emph{Ensemble of different instruments}: a heterogeneous group blends distinct tonal colours and enables richer contrapuntal interaction.  
    (d) \emph{Culture-specific ensemble}: a traditional instrumental configuration (e.g.\ guzheng quartet, gamelan group) that captures the performance idioms of a particular musical culture.}
\label{fig:music-scene-examples}
\end{figure}

Figure~\ref{fig:music-scene-examples} presents examples for each of the scene types defined in Section~\ref{sec:background}. These examples further underscore the performance diversity that Music AVQA methods must accommodate, ranging from sparse solo stages to densely populated, culturally nuanced ensembles.

\begin{figure}[ht]
    \centering
    \subfigure[Methods on Music-AVQA~\cite{9879157}.]{
        \includegraphics[width=0.99\linewidth]{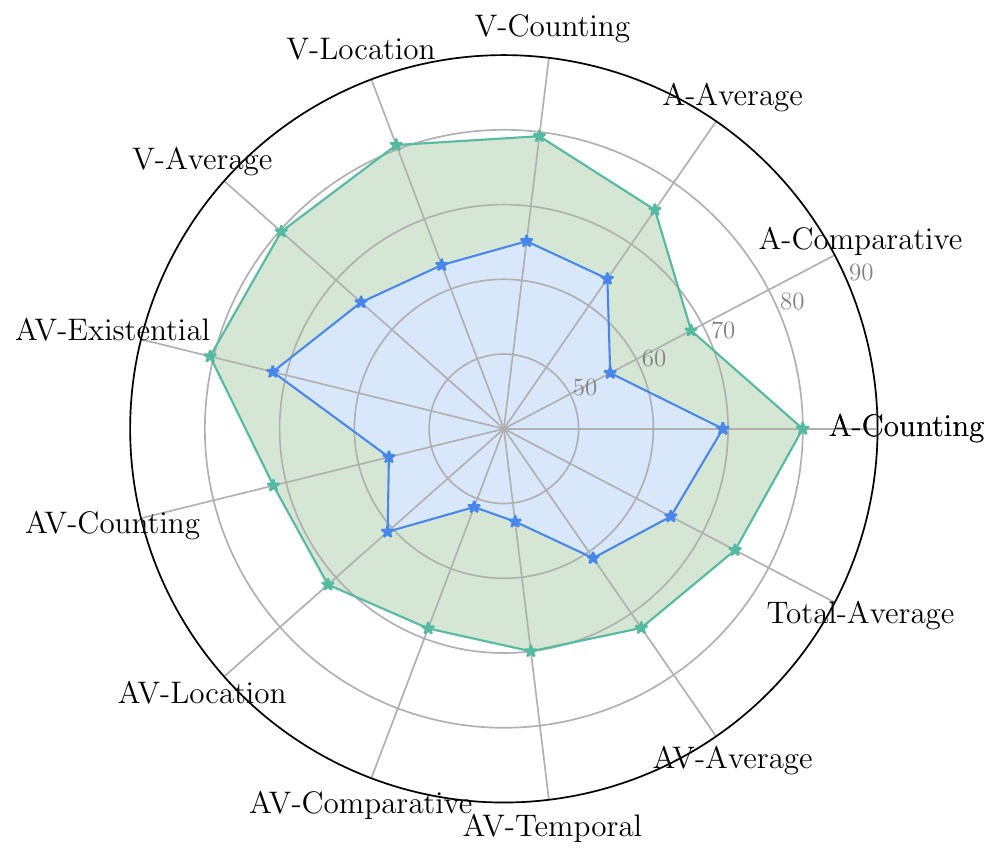}
        \label{subfig:music-avqa-radar-big}
    }

    \subfigure[Methods on Music-AVQA-R~\cite{ma2024look}.]{
        \includegraphics[width=0.99\linewidth]{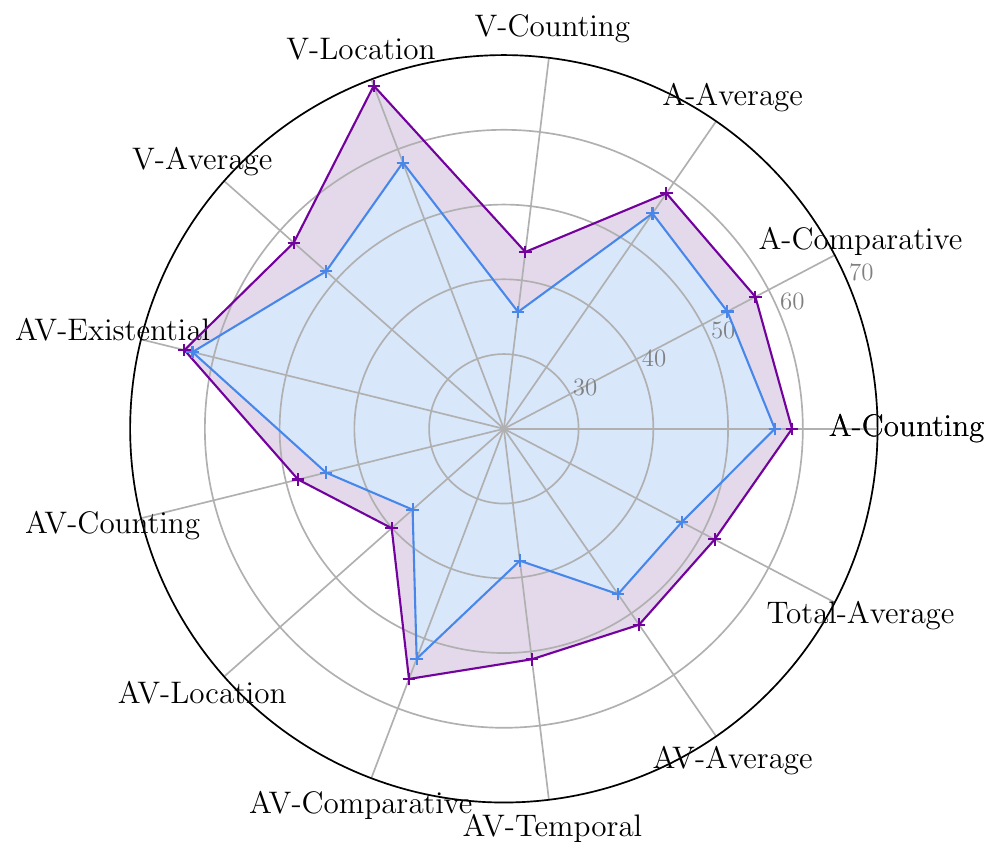}
        \label{subfig:music-avqa-r-radar-big}
    }
\caption{
Radar plots showing the per-type average accuracy of model groups with and without \textbf{spatial-temporal design} across 13 QA categories on (a) Music-AVQA~\cite{9879157} and (b) Music-AVQA-R~\cite{ma2024look}. Each axis corresponds to a QA type spanning audio, visual, and audio-visual reasoning, including the overall average (Total-Average). The filled ~\colorbox[HTML]{d5e7d4}{green} polygon in Figure \ref{subfig:music-avqa-radar} and ~\colorbox[HTML]{e4d9ea}{purple} polygon in Figure \ref{subfig:music-avqa-r-radar} represent the mean accuracy across QA types for models with spatial-temporal design, while the~\colorbox[HTML]{d9e7fb}{blue} polygon represents the average performance of models without such design. Models with spatial-temporal design consistently achieve higher accuracy across all modality groups. These advantages persist under distribution shift in the robustness-focused Music-AVQA-R dataset.}
\label{fig:radar-big}
\end{figure}

% \clearpage
\section{Representative Examples of Music AVQA Question Types}
\label{appendix:avqaTypeEg} 

\begin{figure*}[ht]
    \centering
    \subfigure[Example of Audio Counting QA.]{%
        \includegraphics[width=0.48\linewidth]{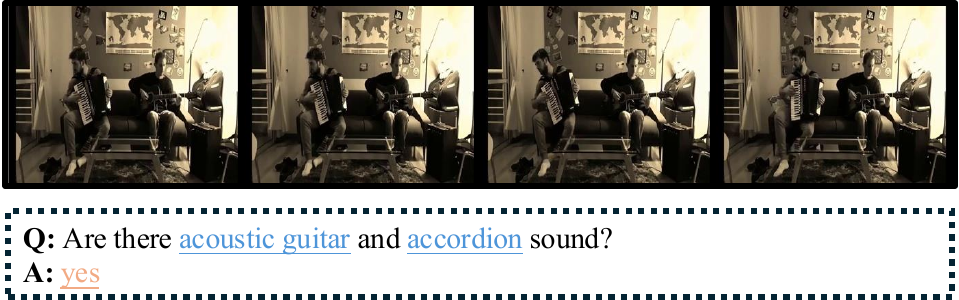}
        \label{subfig:task-a-counting}
    }
    \subfigure[Example of Audio Comparative QA.]{%
        \includegraphics[width=0.48\linewidth]{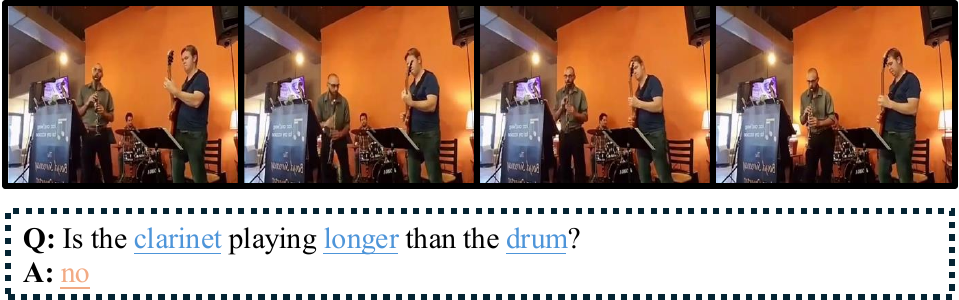}
        \label{subfig:task-a-comp}
    }
    \subfigure[Example of Visual Counting QA.]{%
        \includegraphics[width=0.48\linewidth]{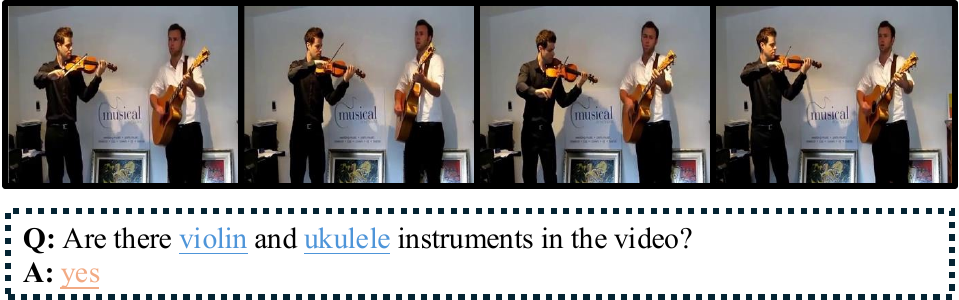}
        \label{subfig:task-v-counting}
    }
    \subfigure[Example of Visual Localization QA.]{%
        \includegraphics[width=0.48\linewidth]{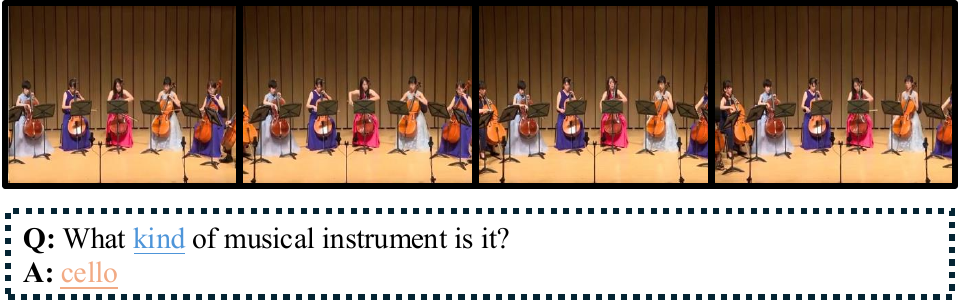}
        \label{subfig:task-v-location}
    }
    \subfigure[Example of Audio-Visual Existential QA.]{%
        \includegraphics[width=0.48\linewidth]{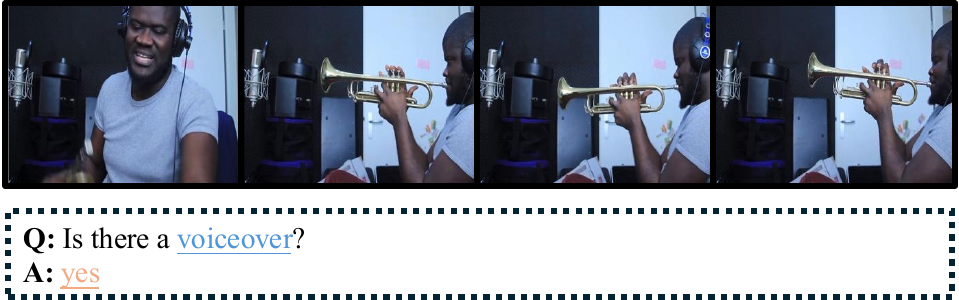}
        \label{subfig:task-av-ext}
    }
    \subfigure[Example of Audio-Visual Counting QA.]{%
        \includegraphics[width=0.48\linewidth]{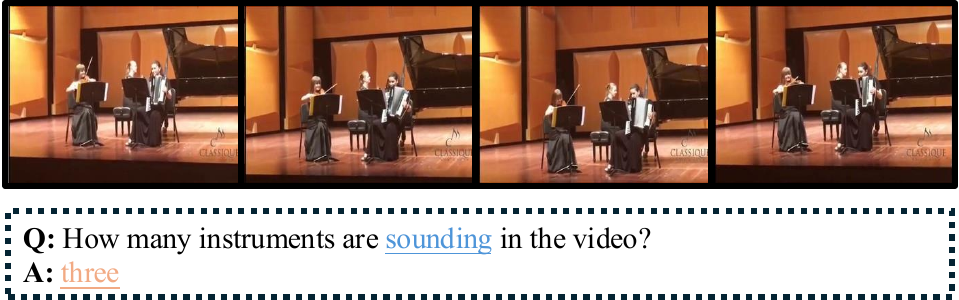}
        \label{subfig:task-av-counting}
    }
    \subfigure[Example of Audio-Visual Localization QA.]{%
        \includegraphics[width=0.48\linewidth]{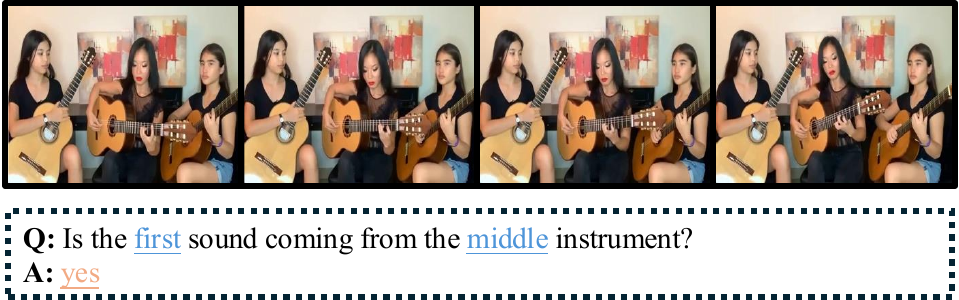}
        \label{subfig:task-av-local}
    }
    \subfigure[Example of Audio-Visual Comparative QA.]{%
        \includegraphics[width=0.48\linewidth]{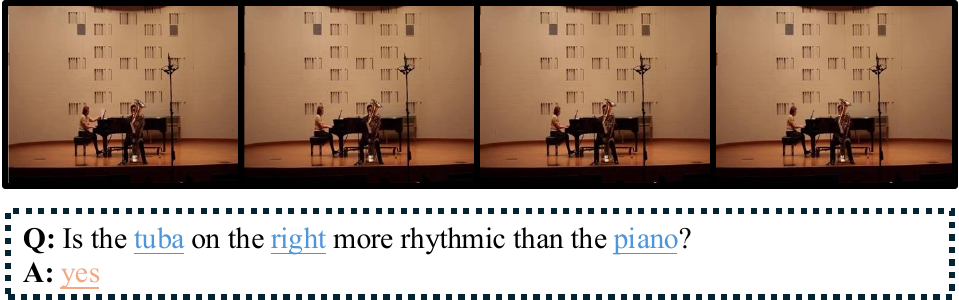}
        \label{subfig:task-av-comp}
    }
    \subfigure[Example of Audio-Visual Temporal QA.]{%
        \includegraphics[width=0.48\linewidth]{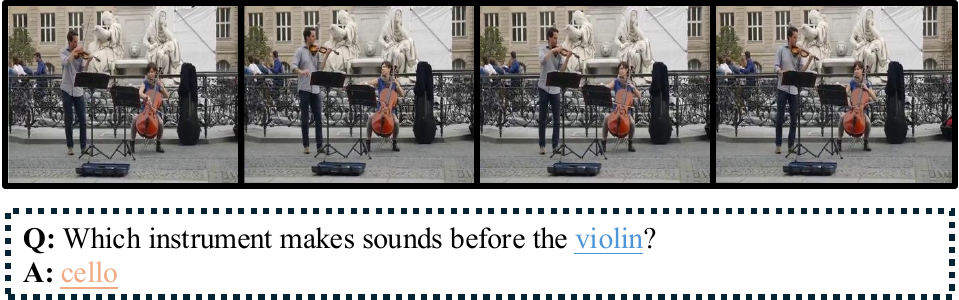}
        \label{subfig:task-av-temp}
    }
    % \vspace{-8pt}
\caption{
Examples of Music AVQA question types spanning audio, visual, and audio-visual modalities, including counting, comparison, localization, existential, and temporal QA.
}
\label{fig:samples-tasks}
\end{figure*}

Table~\ref{tab:qa-examples} and Figure~\ref{fig:samples-tasks} provide representative examples of the Music AVQA question types. These illustrate how each type manifests itself across audio, visual and audio-visual modalities, highlighting the multimodal and fine-grained nature of the task.

\begin{table*}[htbp]
\centering
\caption{Examples of questions in Music AVQA categorized by modality involved and task type.}
\label{tab:qa-examples}
\footnotesize
\resizebox{1.0\textwidth}{!}{
\begin{tabular}{p{2.5cm}p{2cm}p{6.8cm}p{1.4cm}}
\toprule
\textbf{Modality} & \textbf{Task Type} & \textbf{Question} & \textbf{Answer} \\
\midrule
\multirow{2}{*}{Audio} 
& Counting &Are there acoustic guitar and accordion sound?& Yes \\
& Comparative & Is the clarinet playing longer than the drum? & No \\

\midrule
\multirow{2}{*}{Visual} 
& Counting &Are there violin and ukulele instruments in the video?& Yes \\
& Localization & What kind of musical instrument is it? & Cello \\

\midrule
\multirow{5}{*}{Audio Visual} 
& Existential & Is there a voiceover? & Yes \\
& Counting & How many instruments are sounding in the video? & Three \\
& Localization & Is the first sound coming from the middle instrument? & Yes \\
& Comparative & Is the tuba on the right more rhythmic than the piano? & Yes \\
& Temporal & Which instrument makes sounds before the violin? & Cello \\
\bottomrule
\end{tabular}
}
\end{table*}

\section{Full-Size Versions of Figures 2 and 3}

\begin{figure*}[ht]
    \centering
    \subfigure[Audio Counting QA.]{%
        \includegraphics[width=0.32\linewidth]{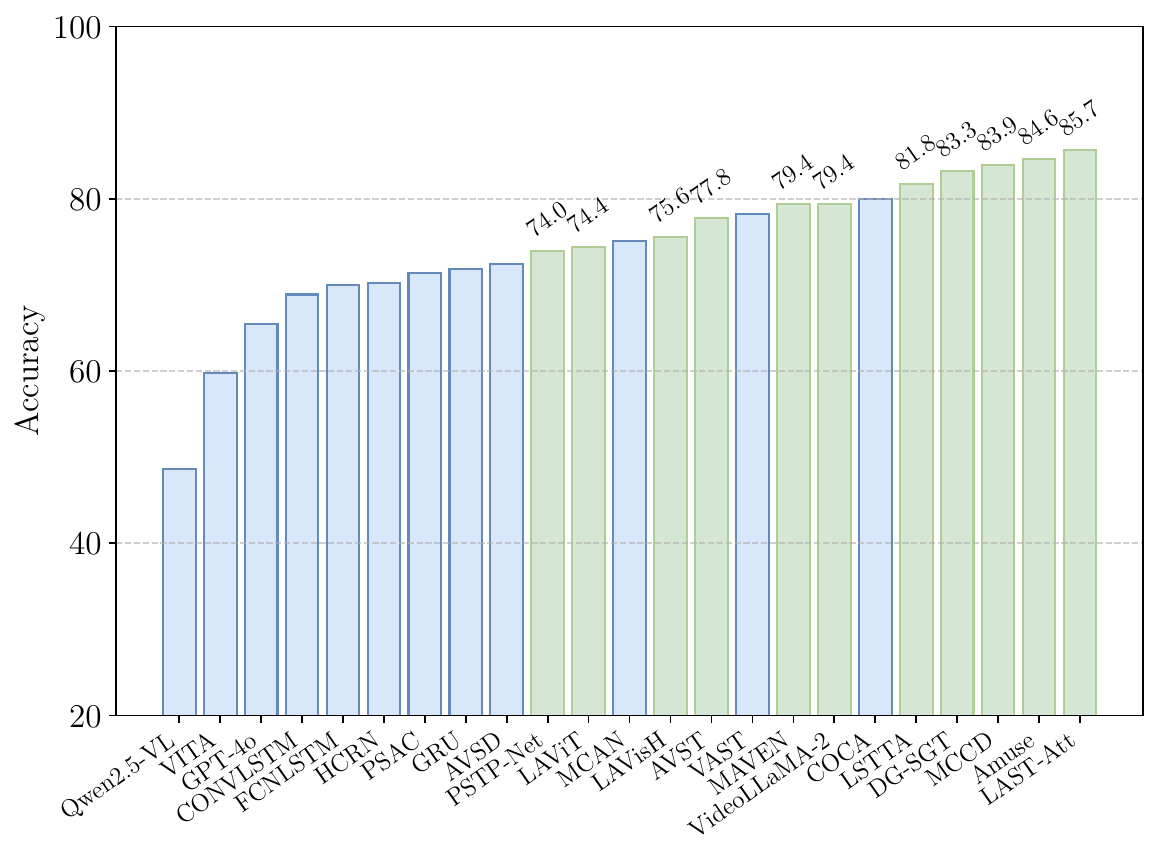}
        \label{subfig:a-counting-big}
    }
    \subfigure[Audio QA Average.]{%
        \includegraphics[width=0.32\linewidth]{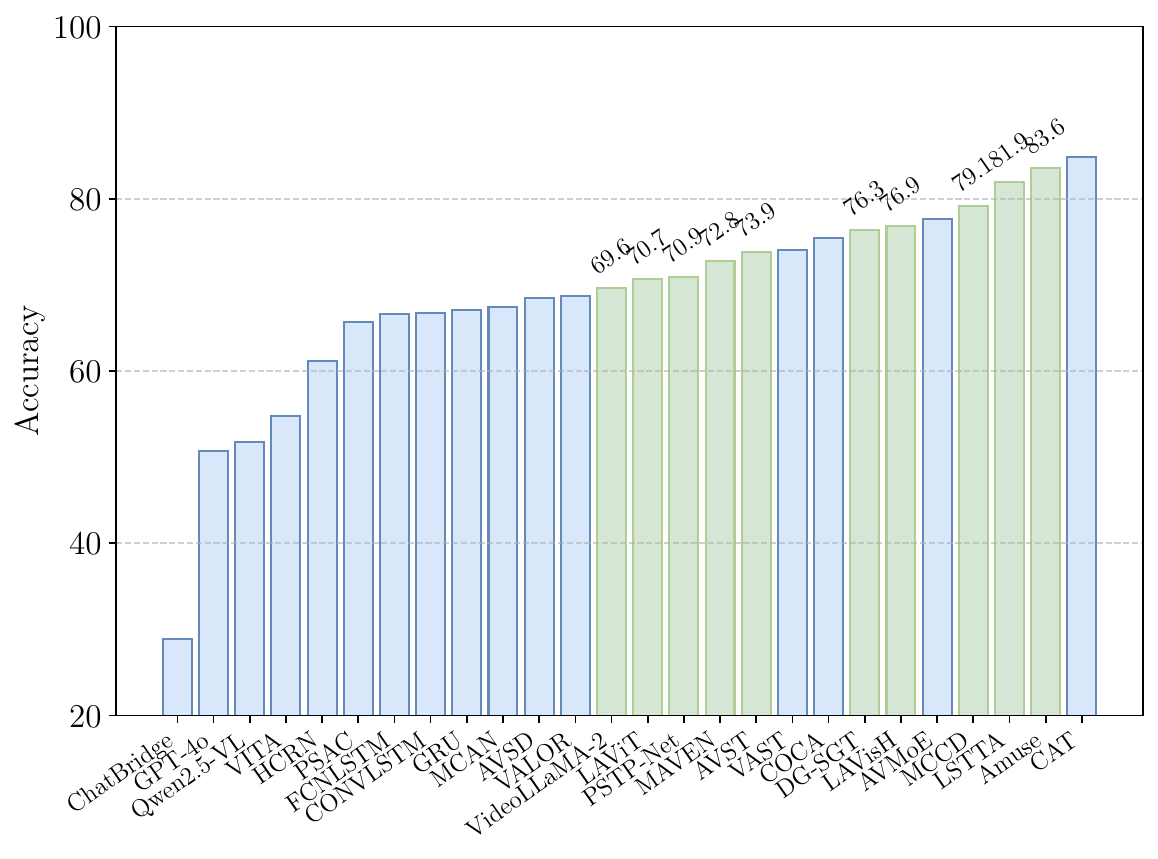}
        \label{subfig:a-average-big}
    }
    \subfigure[Visual Counting QA.]{%
        \includegraphics[width=0.32\linewidth]{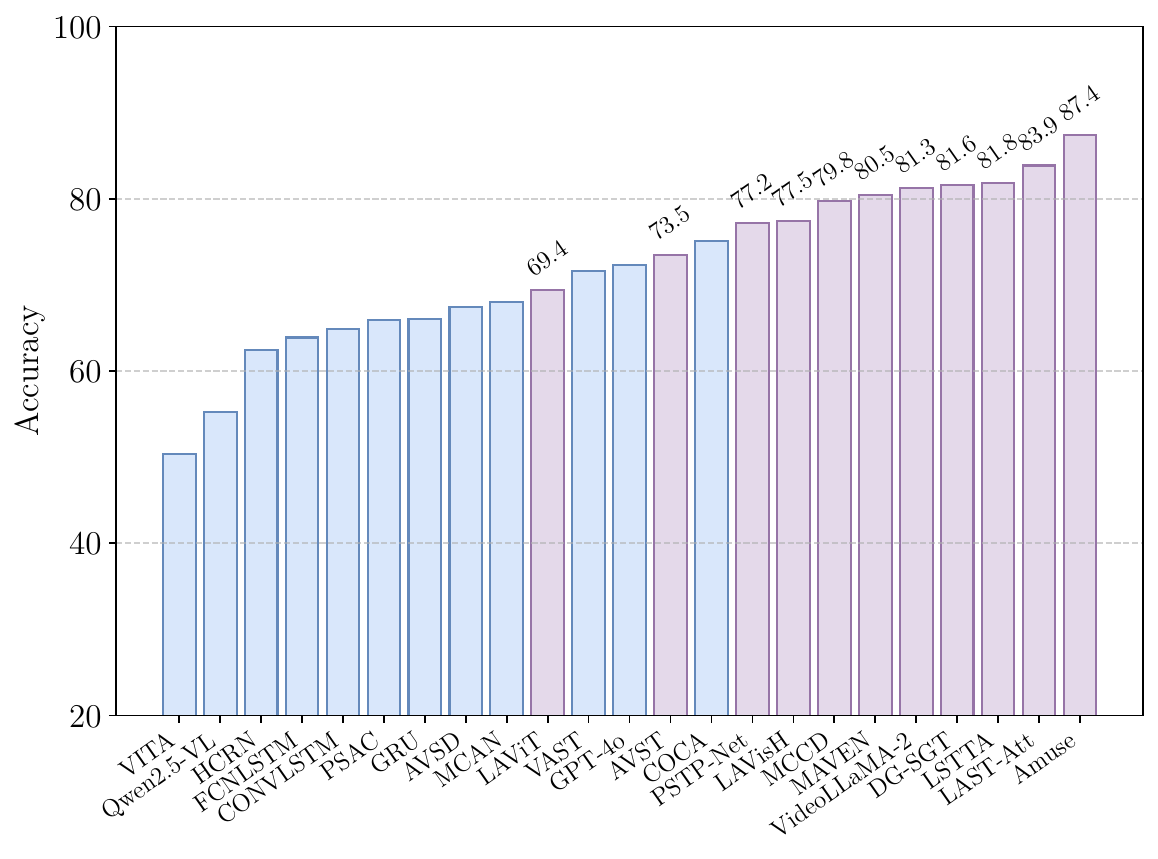}
        \label{subfig:v-counting-big}
    }\\[0em] 

    \subfigure[Visual Location QA.]{%
        \includegraphics[width=0.32\linewidth]{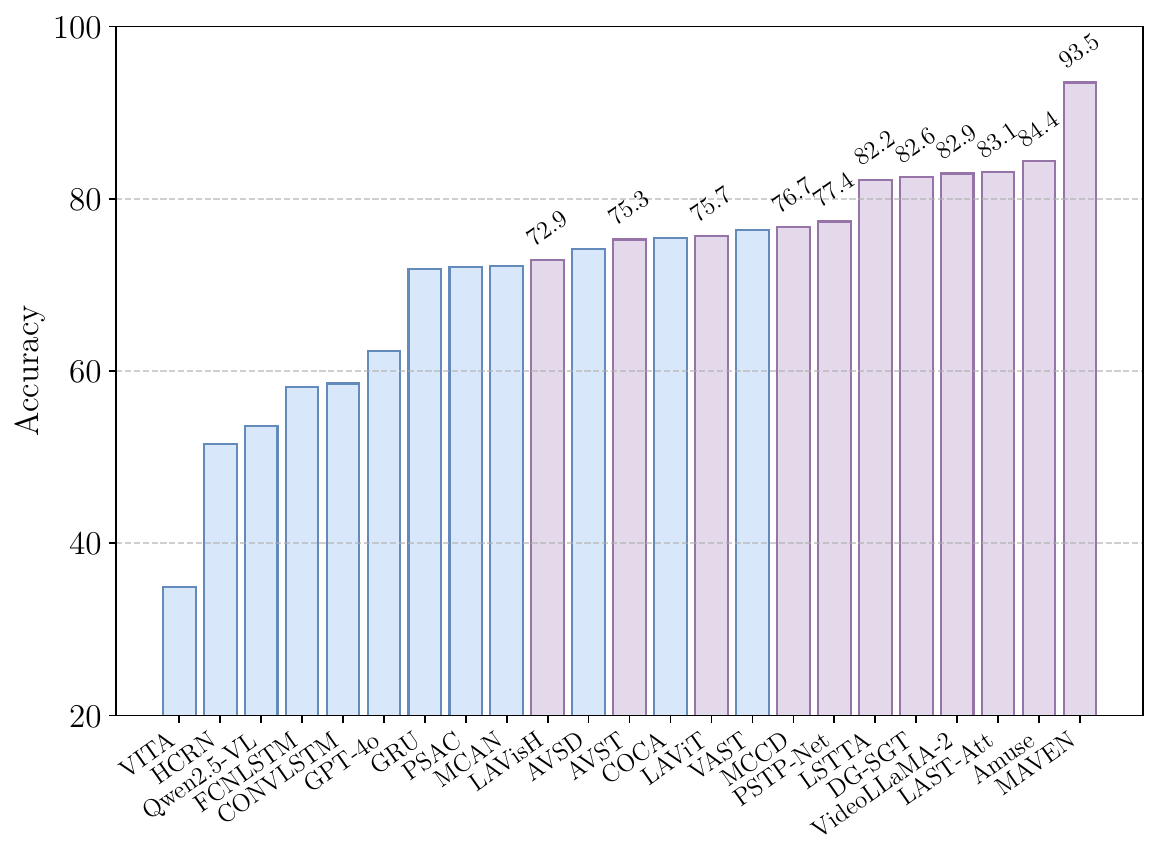}
        \label{subfig:v-location-big}
    }
    \subfigure[Visual QA Average.]{%
        \includegraphics[width=0.32\linewidth]{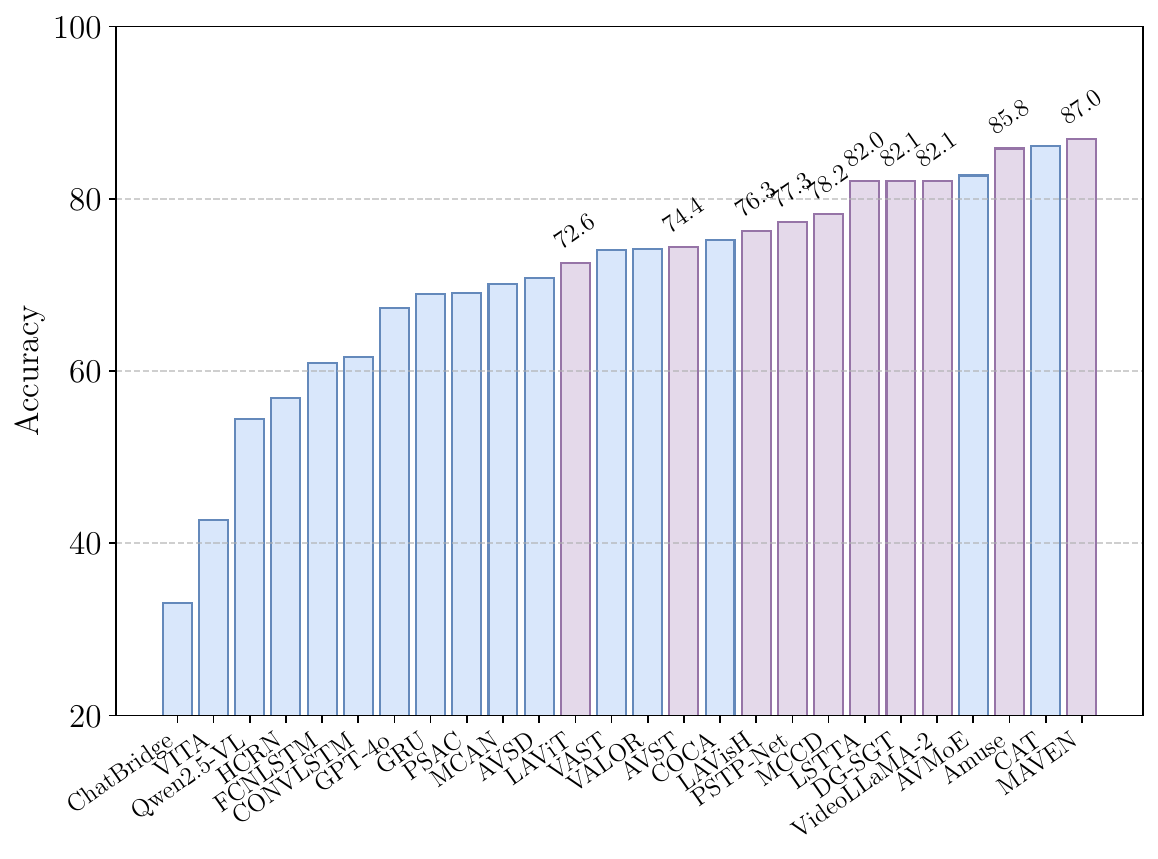}
        \label{subfig:v-average-big}
    }
    \subfigure[Audio-Visual Location QA.]{%
        \includegraphics[width=0.32\linewidth]{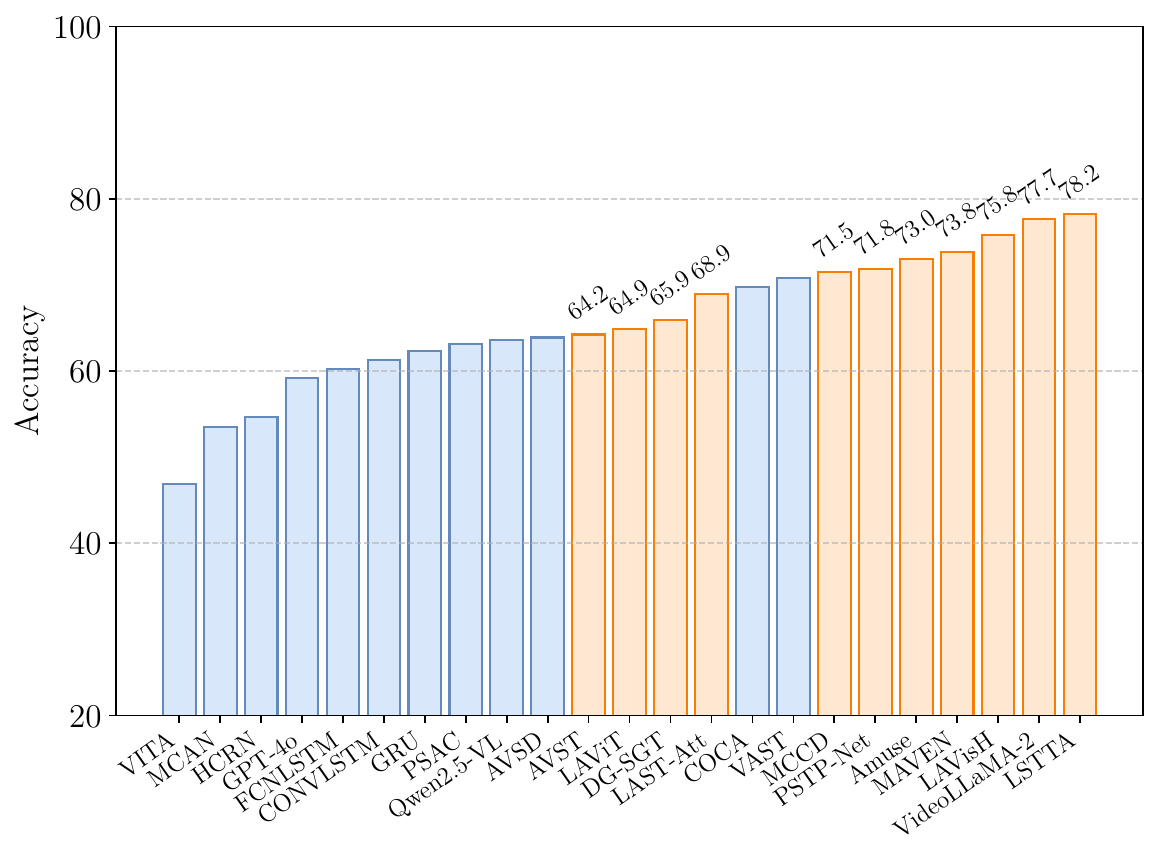}
        \label{subfig:av-location-big}
    }\\[0em]

    \subfigure[Audio-Visual Comparative QA.]{%
        \includegraphics[width=0.32\linewidth]{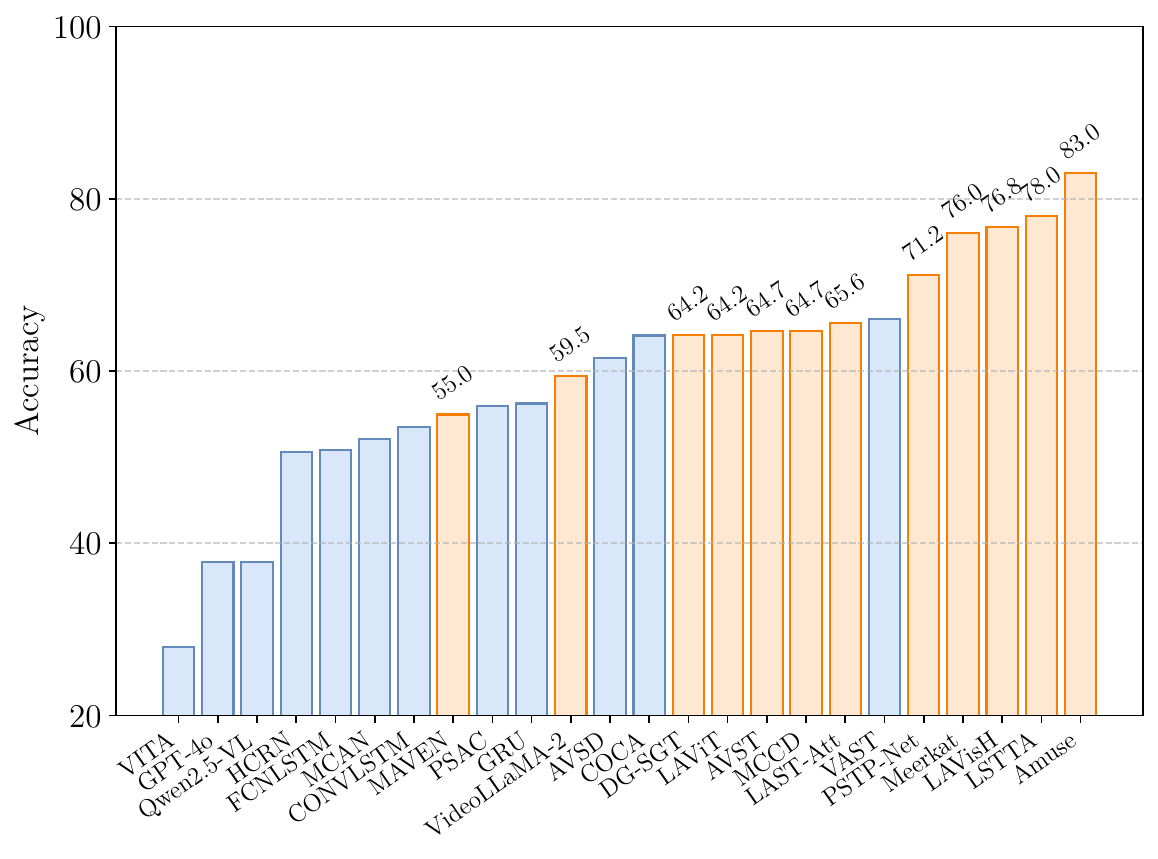}
        \label{subfig:av-comparative-big}
    }
    \subfigure[Audio-Visual Temporal QA.]{%
        \includegraphics[width=0.32\linewidth]{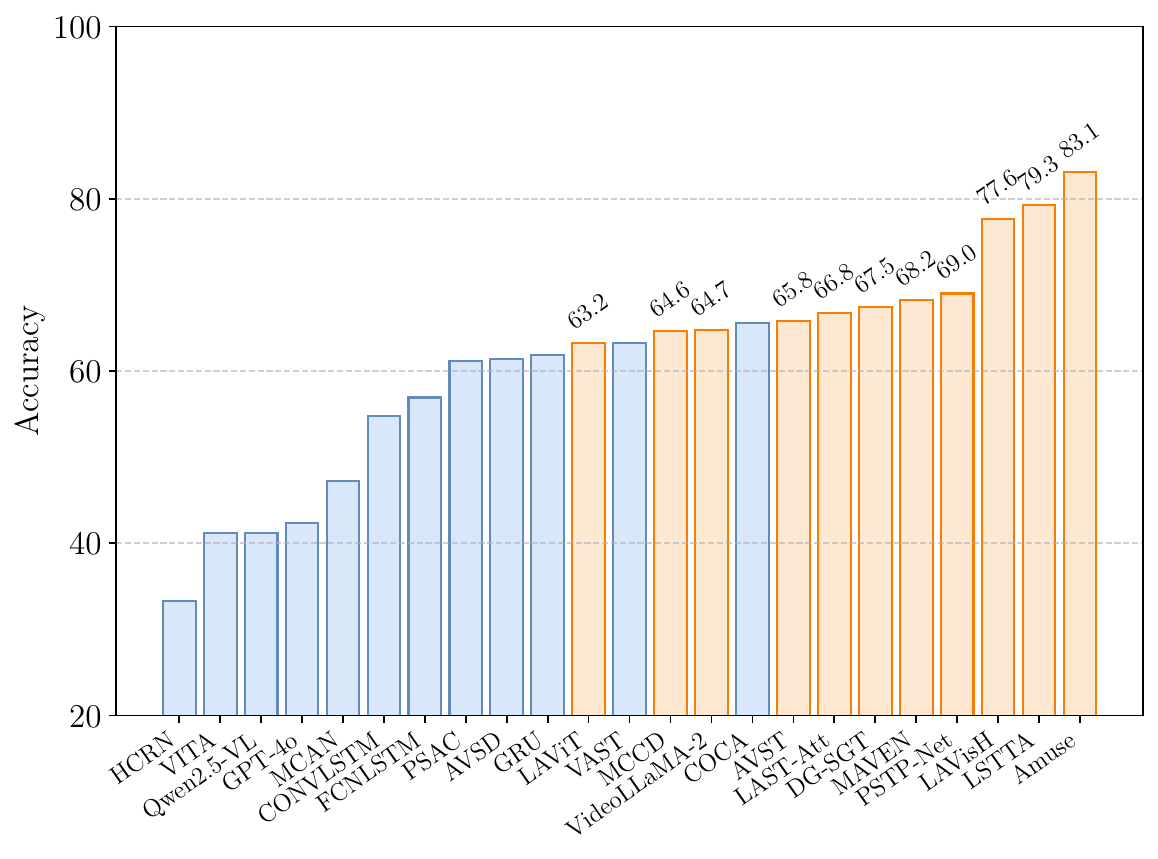}
        \label{subfig:av-temporal-big}
    }
    \subfigure[Audio-Visual QA Average.]{%
        \includegraphics[width=0.32\linewidth]{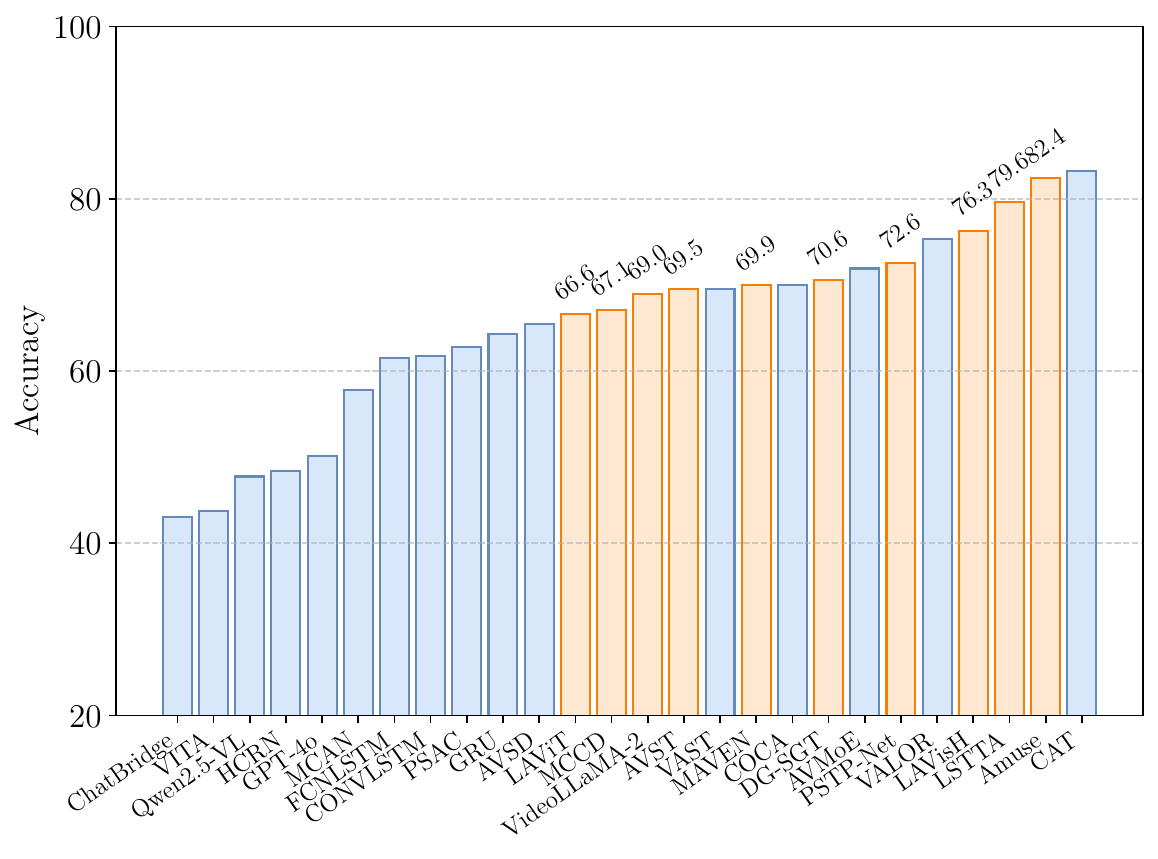}
        \label{subfig:av-average-big}
    }
    % \vspace{-8pt}
\caption{
Accuracy comparison of Music AVQA models across representative question types, grouped by modality: (a–b) Audio, (c–e) Visual, and (f–i) Audio-Visual. Each bar corresponds to a model and is color-coded based on whether it incorporates \textbf{spatial-temporal design} for the relevant task type: bars in ~\colorbox[HTML]{d5e7d4}{green}, ~\colorbox[HTML]{e4d9ea}{purple}, and ~\colorbox[HTML]{ffe8d2}{orange} represent models that apply spatial-temporal modeling to Audio-related, Visual-related, and Audio-Visual-related question answering, respectively; bars in ~\colorbox[HTML]{d9e7fb}{blue} represent models without spatial-temporal design. Across most categories, models with spatial-temporal components tend to perform more accurately, particularly on tasks requiring temporal reasoning or spatial localization. These patterns suggest that incorporating spatial-temporal design supports more effective reasoning in musically structured multimodal environments.
}
\label{fig:spatial-analysis-big}
\end{figure*}

This Appendix Section presents full-size reproductions of the main-text figures. 
Specifically, Figure~\ref{fig:spatial-analysis-big} is the enlarged version of Figure~\ref{fig:spatial-analysis} (Figure 2, accuracy comparisons by modality), and Figure~\ref{fig:radar-big} is the enlarged version of Figure~\ref{fig:radar} (Figure 3, per-type radar plots). The scale is increased to improve legibility of axes, legends, and model names.

\label{sec:clr-visual} 

\section{Details of Music AVQA Datasets}
\label{sec:music_dataset}
We summarize the three publicly released Music-AVQA benchmarks and their successive extensions (\S~\ref{appendix:music-avqa-datasets}), then contrasts Music-AVQA with several general-purpose AVQA datasets to underline the domain-specific challenges posed by musical performance videos (\S~\ref{appendix:other-datasets}).

\subsection{Music AVQA Datasets}\label{appendix:music-avqa-datasets}
Table~\ref{table:dataset} provides a summary of three representative datasets specifically designed for Music Performance Audio-Visual Question Answering (Music AVQA) tasks.
\setlength{\tabcolsep}{1pt}

\begin{table}[htbp]
    \centering
\caption{Evolution and characteristics of Music AVQA datasets: a comparative overview of MUSIC-AVQA~\cite{9879157}, MUSIC-AVQA v2.0~\cite{10484352}, and MUSIC-AVQA-R~\cite{ma2024look} benchmarks.}
    \footnotesize
    \begin{tabular}{p{0.3\linewidth}|p{0.7\linewidth}}
    \toprule
    \textbf{Dataset} & \textbf{Brief Description} \\ 
    \midrule
    MUSIC-AVQA \cite{9879157} & The MUSIC-AVQA dataset represents a significant contribution to audio-visual question answering research, comprising 9,288 videos with over 150 hours of musical performances covering 22 instruments, generating 45,867 question-answer pairs. The dataset is randomly split into training, validation, and testing sets with 32,087, 4,595, and 9,185 QA pairs respectively, spanning 33 question templates across 9 question types including existential, location, counting, comparative, and temporal questions.\\
    \hline
    MUSIC-AVQA v2.0 \cite{10484352} &  The MUSIC-AVQA v2.0 dataset builds upon the original MUSIC-AVQA by addressing data bias issues, comprising 10,518 videos (9,288 from the original plus 1,230 new videos) with musical performances covering 22 instruments, generating approximately 54,000 question-answer pairs. The balanced dataset splits into training, validation, and testing sets with 36,700, 5,250, and 10,819 QA pairs respectively, spanning 33 question templates across 9 question types. The authors specifically balance 15 biased templates by ensuring no dominant answers exceed 60\% for binary questions or 50\% for multi-class questions, particularly enhancing representation of underrepresented answers in existential, counting, temporal, location, and comparative question categories.\\
    \hline
    MUSIC-AVQA-R \cite{ma2024look} & The MUSIC-AVQA-R dataset proposed in this paper is an extension of MUSIC-AVQA specifically designed to evaluate the robustness of audio-visual question answering models. It expands the original test set through a human-machine collaboration mechanism that rephrases each question 25 times, increasing the number of questions from 9,129 to 211,572, and introduces distribution shifts to categorize questions into head (common) and tail (rare) samples. Compared to the original dataset, MUSIC-AVQA-R features a vocabulary size of 465 (five times larger than MUSIC-AVQA), provides more diverse question formulations while preserving inherent biases in the training and validation sets, and offers three evaluation metrics—head accuracy, tail accuracy, and overall accuracy—enabling researchers to assess model performance in both in-distribution and out-of-distribution scenarios, making it the first dataset specifically designed for robustness evaluation in audio-visual question answering tasks. \\
    \bottomrule
    \end{tabular}
    \label{table:dataset}
\end{table}

\subsection{Key Difference from Other AVQA Datasets}\label{appendix:other-datasets}
Table~\ref{tab:other_datasets} contrasts the Music-AVQA dataset \cite{9879157} with several widely-used AVQA benchmarks \cite{yang2022avqa, mangalam2023egoschema, xie2024funqa, 10.1109/TPAMI.2024.3479776, 9053174}. For each dataset, we highlight the most salient divergence from the music-specific setting, focusing on aspects such as task format, content domain, temporal scope, and the presence or absence of fine-grained musical reasoning.

\setlength{\tabcolsep}{1pt}
\begin{table}[htbp]
    \centering
\caption{Other representative benchmarks (AVQA~\cite{yang2022avqa}, EgoSchema~\cite{mangalam2023egoschema}, FunQA~\cite{xie2024funqa}, VALOR-1M~\cite{10.1109/TPAMI.2024.3479776}, and VGG-Sound~\cite{9053174}) and the key difference each bears with respect to Music-AVQA~\cite{9879157}.}
    \footnotesize
    \begin{tabular}{p{0.3\linewidth}|p{0.7\linewidth}}
    \toprule
    \textbf{Dataset} & \textbf{Key Difference} \\ 
    \midrule
    AVQA~\cite{yang2022avqa}&Builds multiple-choice QA on everyday VGG-Sound clips; questions target generic activities and causal relations in real-life videos, so it lacks the fine-grained instrument/sound localization and music-theory knowledge required in MUSIC-AVQA.\\
    \hline
    EgoSchema~\cite{mangalam2023egoschema}&Uses first-person (Ego4D) footage that is three-minutes long, stressing long-range temporal reasoning in egocentric daily tasks; audio cues are incidental and the task is 5-way multiple choice, very different from the short, professionally filmed music performances and open-ended answers in MUSIC-AVQA.\\
    \hline
    FunQA~\cite{xie2024funqa}&Focuses on``surprising'' humour/creative/magic clips (4.3 k videos, 312 k QAs) that test commonsense violations; audio is often background and questions centre on counter-intuitive visual events, not on synchronised musical notes or instrument semantics.\\
    \hline
    VALOR-1M~\cite{10.1109/TPAMI.2024.3479776}&A pre-training corpus (1 M videos) with tri-modal captions meant for retrieval/captioning; QA supervision is extremely sparse and relies on auto-generated subtitles, so it serves as a foundation model resource rather than a targeted AVQA evaluation set like MUSIC-AVQA.\\
    \hline
    VGG-Sound~\cite{9053174}&It is an audio-visual correspondent dataset consisting of short clips of audio sounds from YouTube. And it provides raw audio–visual correspondence but no question–answer supervision or fine-grained reasoning labels.\\
    \bottomrule
    \end{tabular}
    \label{tab:other_datasets}
\end{table}

\clearpage
\section{How Music AVQA Differs from Traditional Multimodal Tasks}
\label{sec:difftasks} 
Recent progress in multimodal research has greatly expanded the scope of multimodal modeling, spanning vision--language pretraining~\cite{zhang2025pretrained,zhang2022look,li2025openvision,liu2025openvision,chen2024taskclip}, video captioning~\cite{jian2023bootstrapping}, multimodal QA~\cite{li2024towards,diao2025protovqa}, multimodal reasoning~\cite{han2024infimm,xie2026q,diao2026addressing,wang2025consistency,wang2025fakesv,wang2025care,wang2024eaco,zhang2025overcoming}, multimodal content generation and dialogue~\cite{gao2024gem,he2025intentenhan,diao2025ft2tf,jian2024expedited}, retrieval~\cite{he2025enhancing,liu2026cast}, audio--visual representation learning~\cite{gong2022contrastive,diao2023av}, multimodal classification and clustering~\cite{qian2022co,yao2024multi,yao2024customized,xie2024uncertainty,xie2025multi}, fMRI understanding and decoding~\cite{wei2025more,wei2025fmri}. Against this broader backdrop, Music AVQA stands out within audio–visual–text (AVT) tasks for requiring models to disentangle \emph{polyphonic, continuous} sound streams, bind them to precise visual sources, and reason with explicit musical knowledge. Below, we discuss several representative multimodal task settings as points of comparison.

\paragraph{Generic Audio–Visual Question Answering.} Classic AVQA benchmarks~\cite{45857,9053174,tian2018audio,zhouavsbench} present short clips with a single audible event (e.g., a dog bark) and ask coarse ``what/where/when'' questions that can be answered once the sounding object is localized and its time span identified. The AVQA dataset itself illustrates this design: most videos contain one foreground sound and minimal polyphony, so millisecond-level alignment is unnecessary~\cite{9879157,yang2022avqa}. Music AVQA, in contrast, poses queries such as ``Which violin enters after the flute?'', which requires the model to track multiple \emph{overlapping} instruments and reason over precise temporal order, a level of granularity generic AVQA never targets.

\paragraph{Video Captioning.} Datasets like MSR-VTT~\cite{xu2016msr}, GIF~\cite{li2016tgif}, ActivityNet Captions~\cite{krishna2017dense}, How2~\cite{sanabria2018how2}, and Vatex~\cite{wang2019vatex} evaluate whether a system can produce one or two fluent sentences that summarize the \emph{gist} of a clip. Moreover, temporal mis-alignment of a few seconds or the omission of background sounds rarely affects the score. Music AVQA eliminates summarisation entirely and replaces it with beat-accurate, question-driven reasoning: the model must pinpoint note onsets, match them to performers, and compare rhythmic or dynamic patterns tasks far beyond the scope of generic captioning.

\paragraph{Instructional Video QA.} TutorialVQA~\cite{colas-etal-2020-tutorialvqa}, YouCookQA~\cite{9423019} and HowToVQA69M~\cite{yang2021just} frame question answering around narrated procedural videos whose audio channel is dominated by speech that explicitly describes each step. The narration acts as a guide-track and overlapping non-speech sounds are rare and not queried. Music AVQA removes narration altogether and treats the dense musical audio as the primary reasoning target, forcing models to infer structure (beats, phrases, instrument entrances) directly from raw sound rather than from textual guidance.

\paragraph{Audio–Visual Scene-Aware Dialogue.} 
The AVSD dataset~\cite{alamri2019audio} and its follow-up~\cite{alamri2018audio,schwartz2019simple,d2020overview,shah2022audio} train systems to hold multi-turn conversations about short ($\approx$ 10 s) household videos, grounding answers in coarse scene context while maintaining dialog coherence. Acoustic events are typically brief (speech, clatter) and alignment at $\pm 1$ s suffices. Music AVQA disregards conversational flow and instead demands hierarchical timing: beats, bars, sections. Every answer depends on tight audio–video synchrony, not on dialog history management. Recent studies on diversified and persona-aware dialogue generation~\cite{li2023bilateral,li2024distinct} further underscore the importance of fine-grained temporal and semantic alignment—even in unimodal dialogue settings.

\paragraph{Multimodal Sentiment Analysis.} Benchmarks such as CMU-MOSI~\cite{zadeh2016mosi} and CMU-MOSEI~\cite{tsai-etal-2019-multimodal} fuse text transcripts, facial expression and prosody to predict sentiment or emotion over 5 to 30 second clips. Overlap between speakers or sound sources is noise to suppress, and no explicit source-binding is required. In Music AVQA overlap is the \emph{signal}: models must isolate each instrument’s contribution, bind it to its on-screen performer and reason about their relationships (loudness, order, count). Other emotion datasets, like IEMOCAP~\cite{busso2008iemocap}, MELD~\cite{poria-etal-2019-meld}, and CH-SIMS~\cite{yu-etal-2020-ch}, use 5–30 s clips, treat overlapping voices as background noise, and label sentiment at the utterance—rather than source—level.

\paragraph{Cross-Modal Retrieval.} Contrastive systems such as CLIP~\cite{radford2021learning} (vision–text), ALIGN~\cite{jia2021scaling}, LiT~\cite{zhai2022lit}, or ImageBind~\cite{girdhar2023imagebind} (audio–vision–text) learn global embeddings and judge success by top-$k$ similarity—minor temporal or spatial errors barely change the score. Music AVQA instead penalises any mis-alignment: swapping left/right instruments or missing a single beat flips the answer. The task therefore demands persistent token-level grounding rather than coarse embedding proximity.

Where existing AVT tasks rely on sparse events, speech cues or global embeddings, Music AVQA alone couples \emph{dense polyphonic audio}, \emph{frame-accurate audio–visual alignment}, and \emph{music-theoretic knowledge}. It thus sets a much higher bar for multimodal reasoning than traditional benchmarks.

\section{Details of Music AVQA Methods with Spatial-Temporal Designs}
\label{baseline}

Table~\ref{table:S-T_design_specifics} illustrates methods incorporating explicit \textbf{spatial-temporal} design components in detailed. 

\onecolumn
\setlength{\tabcolsep}{1pt}
\begin{longtable}{%
    >{\raggedright\arraybackslash}p{0.24\linewidth}|
    >{\raggedright\arraybackslash}p{0.74\linewidth}}
    \caption{Description of representative methods for \textbf{spatial-temporal design} for Music AVQA.}
\label{table:S-T_design_specifics}\\
\toprule
\textbf{Paper/Work} & \textbf{Brief Description} \\
\midrule
\endfirsthead         
\toprule
\textbf{Paper / Work} & \textbf{Brief Description} \\
% \midrule
\endhead            
\midrule
\multicolumn{2}{r}{\footnotesize\textit{Continued on next page}}\\
\endfoot
\bottomrule
\endlastfoot
\textsc{Amuse}~\cite{diao2024learning}&Focuses on music performance scenarios by aligning time segments in both audio and video streams via a cross-attention paradigm. Exploits synchronized features (such as beat-level or note-level alignment) to capture subtle temporal dependencies among instruments in dense music passages. By integrating rhythmic cues and cross-modal interactions, it is particularly suited for questions that involve multiple instruments playing simultaneously or changing their patterns over time.  \\
    \hline
    \textsc{AVST}~\cite{9879157}&Proposes a spatio-temporal grounded audio-visual approach that explicitly localizes sounding objects in each frame while applying a question-guided temporal attention mechanism. The model grounds audio-visual events and emphasizes which frames (visual) and which segments (audio) are most relevant for question answering. By combining localized visual features and temporal cues, it captures object interactions over time and can better handle questions involving spatial and temporal reasoning.  \\
    \hline
    \textsc{CIGN}~\cite{mo2023class}&Learns audio-visual class tokens and an Audio-Visual Continual Grouping module that, at every time-step, pulls together frame-level spectrogram features and region features into category-aware clusters. A token-distillation schedule preserves past knowledge while the regrouping logic tracks objects and sounds through the video’s timeline, giving the model temporally consistent, cross-modal semantics for spatial-temporal reasoning.\\
    \hline 
\textsc{DCL}~\cite{lv2023disentangled}&Introduces a Disentangled Counterfactual Learning framework to handle physical audio-visual commonsense reasoning tasks. Decomposes video signals into static (time-invariant) and dynamic (time-varying) factors using a VAE-based encoder, enabling clearer separation of constant background features from changing events. Additionally employs a counterfactual intervention module on the dynamic factors to perform causal reasoning, helping the model answer ``what if'' questions related to temporal and event relationships.  \\
    \hline
    \textsc{DG-SCT}~\cite{duan2023cross}&Introduces a Dual-Guided Spatial-Channel-Temporal (DG-SCT) attention layer that is injected in every frozen audio and visual transformer block. Audio prompts steer visual tokens (and vice-versa) via bidirectional attention that highlights salient spatial regions, discriminative feature channels, and pivotal temporal segments, producing fine-grained spatio-temporal alignments that boost related tasks.\\
    \hline
    \textsc{EEMC}~\cite{10.1007/978-3-031-72904-1_12}&Divides audio/video into 1-s slices and fuses them with text through a Temporal Bi-modal Transformer backed by a cached-memory mechanism that magnifies sudden changes across time. The resulting multimodal cue stream then serves as a cross-attention prompt for the segmentation decoder, enabling precise localisation of objects and events as their spatial footprints and temporal order evolve.\\
    \hline
    \textsc{LAST-Att}~\cite{10484352}&	Tackles audio-visual question answering with a repeated cross-attention strategy. Uses Swin-Transformer-v2 for visual frame features and a specialized Audio Spectrogram Transformer for audio, then merges them based on the question. By repeatedly ``attending'' to the most relevant frames and spectrogram patches, it effectively localizes musical actions (e.g., a pianist’s keystrokes) over time. This design is well suited for intricate temporal queries and locating key audio events in dense musical content.  \\
      \hline
      \textsc{LAVisH}~\cite{lin2023vision}&Adds a lightweight Latent Audio-Visual HYbrid adapter to every layer of a frozen ViT. A compact pool of latent tokens acts as a cross-attention bottleneck, letting audio frames gate visual tokens (and vice-versa) as the video unfolds, so spatial patches and framewise dynamics are fused early while keeping the backbone frozen.\\
      \hline
      \textsc{LAViT}~\cite{yun2021pano}&Targets 360° videos with a transformer that augments each patch by a quaternion-based spherical coordinate and aligns it with audio via joint contrastive objectives. The spherical embedding plus an auxiliary audio-skewness prediction head lets the model reason about where (on the sphere) and when a sound arises, delivering fine-grained spatial-temporal grounding beyond normal FOV clips.\\
      \hline
    \textsc{LSTTA}~\cite{liu2023parameter}&A parameter-efficient transfer learning approach for audio-visual-language tasks that adds dedicated adapter modules while freezing large pretrained backbones. Splits temporal modeling into two scales: a short-term semantic interaction module (for capturing local correlations such as brief instrumental flourishes) and a long-term semantic filtering module (for broader progressions over many frames). This structure helps the model identify when, how, and for how long different instruments contribute, achieving a refined spatio-temporal representation.  \\
    \hline
    \textsc{MAVEN}~\cite{ma2025fortisavqa}&Employs a Multimodal Audio-Visual Epistemic Network that cycles between audio, video and text logits, using debiasing constraints to keep modality-specific and fused predictions consistent over time. The cycle guidance implicitly anchors each question to the correct temporal segments while suppressing spurious correlations.\\
    \hline
    \textsc{MCCD}~\cite{ma2024look}&Introduces a Multifaceted Cycle-Collaborative Debiasing objective: KL penalties enlarge the gap between uni-modal and tri-modal logits at every timestep, then force the three unimodal paths to agree with each other. This temporal-cycle training steers attention toward frames (and sounds) that all modalities truly share, yielding stabler spatial-temporal grounding under distribution shift.\\
    \hline
    \textsc{Meerkat}~\cite{10.1007/978-3-031-73039-9_4}&Employs a two-stage mechanism for fine-grained audio-visual grounding in space and time. First uses an Audio-Visual Optimal Transport (AVOpT) module for fine-grained local alignment between audio features and specific image patches. Next, the Audio-Visual Attention Consistency Enforcement (AVACE) module refines cross-modal attention maps to precisely locate audio sources within bounding boxes, enforcing spatial constraints and ensuring attention is focused on the correct visual objects that correspond to the audio signal.  \\
    \hline
    \textsc{PSTP-Net}~\cite{li2023progressive}&Proposes a Progressive Spatio-Temporal Perception framework for audio-visual QA. Divides the selection of relevant information into three modules: (1) the Temporal Segment Selection Module (TSSM) for picking key time segments pertinent to the question; (2) the Spatial Region Selection Module (SRSM) to identify essential visual patches within those segments; and (3) the Audio-guided Visual Attention Module (AVAM) to align selected visual patches with the audio signals. This stepwise process helps isolate question-relevant data and reduce interference.  \\
    \hline
    \textsc{RefAtomNet}~\cite{peng2024referring}&For referring atomic actions, it runs three streams—visual, text and location-semantic tokens— and merges them through novel cross-stream agent-attention blocks. The location-semantic stream provides per-person bounding-box hints over time, letting the network lock onto the described individual before classifying frame-level atomic actions, thus tightly coupling spatial localisation with temporal action cues.\\
    \hline
    \textsc{VideoLLaMA-2}~\cite{cheng2024videollama}&Builds a video-LLM around a Spatial-Temporal Convolution (STC) connector that first performs per-frame spatial mixing and then downsamples temporally, giving the language model a compact yet order-aware token sequence. A jointly-trained audio branch injects synchronized spectrogram tokens, enabling the model to answer audio-visual questions that hinge on both where events happen on screen and when they unfold.\\
\end{longtable}

\twocolumn
\section{Details of Existing Music AVQA Methods}
\label{sec:existing_methods}
\begin{itemize}[leftmargin=*]
\item{\textsc{AVMoE}~\cite{cheng2025mixtures}: The paper proposes a parameter-efficient transfer learning framework for audio-visual tasks by dynamically integrating intra-modal and inter-modal information through a mixture of experts. The approach introduces unimodal adapters to capture within-modality details and cross-modal adapters to model interactions between audio-visual streams, while a lightweight modality-agnostic router dynamically allocates expert weights based on input characteristics. By combining these components, AVMoE adaptively balances modality-specific and cross-modal features, addressing challenges like missing modalities or noisy inputs, thereby enhancing robustness and performance across diverse audio-visual tasks such as AV localization, segmentation, and question answering without requiring full model fine-tuning.}
\item{\textsc{AVSD}~\cite{schwartz2019simple}: The paper proposes an end-to-end baseline for audio-visual scene-aware dialog to enhance virtual assistants by integrating multimodal signals. The method employs an attention mechanism to differentiate useful signals from distractions, while maintaining spatial features from video frames (VGG19/I3D-Kinetics) to preserve contextual details and temporally subsampling frames to improve efficiency. By fusing attended vectors across audio, video, and text modalities, the approach dynamically focuses on relevant cues during answer generation. This integrated framework addresses challenges in holistic dialog management, leveraging cross-modal interactions to outperform prior methods without relying on rigid pipelines, as demonstrated on the audio-visual scene-aware dataset.}
\item{\textsc{AVST}~\cite{9879157}: The paper proposes a novel approach to Audio-Visual Question Answering (AVQA) by integrating multimodal understanding and spatio-temporal reasoning in dynamic audio-visual scenarios. It introduces the MUSIC-AVQA dataset with 45K QA pairs to benchmark the task, while addressing spatial associations through an attention-based sound source localization module (AV-Loc) to link sounds with visual sources. Temporal grounding (Q-Temp) is achieved by using question features to highlight key timestamps, enabling effective encoding of question-aware audio-visual embeddings. These components are fused to jointly represent spatial and temporal cues, overcoming challenges in cross-modal reasoning and enhancing performance in complex audio-visual scenes without relying on single-modality methods. The integrated framework demonstrates superior scene understanding by leveraging multisensory perception and fine-grained spatio-temporal analysis.}
\item{\textsc{AVSiam}~\cite{lin2024siamese}: The paper proposes an efficient and scalable audio-visual learning framework using a shared vision transformer backbone to unify audio and visual processing. The AVSiam model employs a contrastive audio-visual matching objective with a multi-ratio random masking scheme to enhance representation robustness while enabling larger batch sizes for effective contrastive learning. By sharing parameters across modalities, the approach reduces GPU memory footprint and computational costs compared to dual-backbone methods, while maintaining competitive performance on classification and retrieval tasks. This integrated design addresses scalability challenges and modality-handling flexibility without compromising accuracy.}
\item{\textsc{Amuse}~\cite{diao2024learning}: The paper proposes a framework for music audio-visual question answering that addresses the unique challenges of dense, continuous audio-visual signals in musical performances. To exploit multimodal interconnectivity, it employs cross-modal adapters to facilitate early-stage token interactions between Swin-V2 (video), HTS-Audio (audio), and language transformers, while annotating rhythm and music sources in datasets to explicitly model musical characteristics. For temporal alignment, it designs specialized encoders to link musical signals with time dimensions. This integrated approach overcomes limitations of general-purpose AVQA methods by capturing intricate audio-visual relationships in performances, enhancing accuracy for music-specific questions through rhythm-aware and temporally grounded representations.}
\item{\textsc{Att-BLSTM}~\cite{zhou-etal-2016-attention}: The paper proposes an attention-based bidirectional LSTM network (Att-BLSTM) for relation classification to capture decisive semantic information without relying on lexical resources or NLP systems. The model processes raw text through an embedding layer to generate word vectors, while bidirectional LSTM (BLSTM) layers learn high-level features by incorporating both past and future context. An attention mechanism then assigns weights to key words, merging word-level features into a sentence-level vector for classification. By integrating these components, the approach overcomes limitations of manual feature engineering and dependency on external tools, effectively identifying critical semantic cues across sentence positions to improve relation classification performance.}
\item{\textsc{Audio Flamingo}~\cite{10.5555/3692070.3693076}: The paper proposes Audio Flamingo, a novel audio language model designed to enhance large language models' (LLMs) understanding of non-speech sounds and non-verbal speech through three key innovations. It employs a sliding-window audio feature extractor to preserve temporal information in variable-length audio, while cross-attention mechanisms efficiently fuse audio inputs into the LM to reduce computational overhead. The model leverages a curated heterogeneous dataset and a two-stage training approach (pre-training and supervised fine-tuning) to balance close-ended and open-ended tasks. Additionally, it integrates in-context learning (ICL) and retrieval-augmented generation (RAG) through tailored templates and cross-attention masks, enabling few-shot adaptation without fine-tuning. To support multi-turn dialogues, the model is fine-tuned on GPT-4-generated datasets with correlated context. By combining these techniques, Audio Flamingo addresses challenges in audio feature extraction, heterogeneous data training, task adaptation, and dialogue coherence, achieving state-of-the-art performance across.}
\item{\textsc{CAT}~\cite{ye2024cat}: The paper proposes an enhanced Multimodal Large Language Model (MLLM) to improve question answering in dynamic audio-visual scenarios by addressing ambiguity and localization challenges. Key components include a clue aggregator to dynamically capture question-aware audio-visual features for fine-grained grounding, a mixed training strategy combining video-text and audio-text pairs with a novel AVinstruct dataset to strengthen cross-modal awareness, and an AI-assisted Ambiguity-aware Direct Preference Optimization (ADPO) to retrain the model for precise responses. By integrating these innovations, CAT effectively mitigates ambiguous outputs and enhances audio-visual reasoning, outperforming existing methods in Audio-Visual Question Answering (AVQA) tasks.}
\item{\textsc{CIGN}~\cite{mo2023class}: The paper proposes a novel framework for continual audio-visual learning by disentangling class-aware cross-modal representations to mitigate catastrophic forgetting. It introduces learnable audio-visual class tokens to continually aggregate category-wise features through the Audio-Visual Continual Grouping module, while the Audio-Visual Class Tokens Distillation module preserves knowledge from previous tasks by aligning old and new token distributions. By integrating these components, the approach effectively addresses the challenge of mixed audio semantics and forgetting in sequential tasks, enhancing discriminative feature learning across modalities without relying on single-modality or rehearsal-based methods. The CIGN framework demonstrates superior performance in class-incremental audio-visual scenarios through its ability to maintain compact and disentangled representations.}
\item{\textsc{COCA}~\cite{lao2023coca}: The paper proposes a collaborative causal regularization framework (COCA) to address multi-shortcut biases in Audio-Visual Question Answering (AVQA) by integrating causal intervention and dynamic debiasing. The Bias-centered Causal Regularization (BCR) mitigates specific shortcut biases (Q→G, V\&Q→G, A\&Q→G) through counterfactual interventions to disrupt bias-irrelevant causal effects and factual regularization to maintain semantic consistency, while the Multi-shortcut Collaborative Debiasing (MCD) dynamically adjusts debiasing focus per sample using an entropy-driven metric to balance bias contributions. By jointly addressing uni-modal and joint-modal biases through causal introspection and instance-aware adaptation, COCA enhances multimodal reasoning robustness without over-correcting, achieving state-of-the-art performance on MUSIC-AVQA.}
\item{\textsc{CONVLSTM}~\cite{9145807}: The paper proposes a novel approach to enhance temporal reasoning in Audio Question Answering (AQA) by introducing the Diagnostic Audio Question Answering (DAQA) dataset, which comprises natural sound events and programmatically generated questions to probe temporal reasoning skills, while adapting visual question answering methods to AQA reveals their limitations. To address this, the authors develop Multiple Auxiliary Controllers for Linear Modulation (MALiMo), which extends Feature-wise Linear Modulation (FiLM) by incorporating an additional auxiliary controller to process subsampled audio features, thereby enabling dynamic modulation of convolutional network processing based on both input modalities. This integrated approach improves relational and temporal reasoning by jointly leveraging audio and question inputs, overcoming the shortcomings of existing methods in handling complex temporal dependencies within sound sequences.}
\item{\textsc{ChatBridge}~\cite{zhao2023chatbridge}: The paper proposes a multimodal language model that leverages large language models (LLMs) as a universal interface to bridge diverse modalities through language-paired data. ChatBridge integrates modality-specific encoders and perceiver modules to project embeddings into the LLM's semantic space, enabling cross-modal correlation without requiring all paired data combinations. The model undergoes two-stage training: first aligning modalities with language to emergent multimodal abilities, then instruction-finetuning on the MULTIS dataset to enhance zero-shot task generalization. By using language as a catalyst, ChatBridge addresses the challenge of limited multimodal paired data while achieving strong performance across text, image, video, and audio tasks through unified multimodal reasoning and user intent alignment.}
\item{\textsc{CrossMAE}~\cite{guo2024crossmae}: The paper proposes a region-aware audio-visual pre-training framework to enhance cross-modality interaction and fine-grained alignment by extending masked autoencoders. It introduces Cross-Conditioned Reconstruction to reconstruct input pixels conditioned on cross-modal Attentive Tokens, while Cross-Embedding Reconstruction leverages Learnable Queries with positional cues to guide feature reconstruction between modalities, supplemented by contrastive loss for global alignment. By integrating these components, CrossMAE addresses the limitations of prior global feature-based methods, enabling effective region-level understanding and improving performance in both classification and dense prediction tasks without task-specific fine-tuning.}
\item{\textsc{DCL}~\cite{lv2023disentangled}: The paper proposes a disentangled counterfactual learning approach for physical audiovisual commonsense reasoning to infer objects' physics properties from video and audio inputs. The method decouples videos into static (time-invariant) and dynamic (time-varying) factors through a disentangled sequential encoder (DSE) using a variational autoencoder and contrastive loss to maximize mutual information while minimizing cross-factor interference. It further introduces a counterfactual learning module (CLM) to model physical knowledge relationships among objects by applying counterfactual interventions as confounders to enhance causal reasoning. By integrating DSE's disentangled representations with CLM's causal learning, the approach effectively addresses challenges in extracting implicit physical knowledge from multi-modal data, improving reasoning explainability and performance without relying on mixed feature representations.}
\item{\textsc{DG-SCT}~\cite{duan2023cross}: The paper proposes a novel Dual-Guided Spatial-Channel-Temporal (DG-SCT) attention mechanism to enhance large pre-trained models for audio-visual tasks by dynamically adjusting feature extraction through cross-modal guidance. The DG-SCT mechanism leverages audio and visual modalities as soft prompts to adaptively refine features across spatial, channel, and temporal dimensions, while preserving frozen pre-trained parameters. By integrating trainable cross-modal interaction layers into encoders, the approach emphasizes task-relevant information in each modality, addressing limitations of single-modality pre-training. This bidirectional prompting enables fine-grained feature fusion, improving performance on downstream tasks like AVE, AVVP, AVS, and AVQA without full retraining, while also excelling in few-shot and zero-shot scenarios.}
\item{\textsc{EEMC}~\cite{10.1007/978-3-031-72904-1_12}: The paper proposes a novel task called Reference Audio-Visual Segmentation (Ref-AVS) to segment visual objects using expressions enriched with multimodal audio-visual cues, addressing the limitations of unimodal approaches. It introduces the Ref-AVS benchmark with pixel-level annotations and diverse multimodal-cue expressions to enable training and evaluation, while an end-to-end framework leverages a crossmodal transformer to process and integrate multimodal cues for precise segmentation. By simultaneously utilizing audio and visual descriptions in natural language, the approach overcomes challenges in locating objects in dynamic audio-visual scenes, enhancing segmentation accuracy in complex real-world scenarios without relying on manual mask annotations or single-modality references.}
\item{\textsc{FCNLSTM}~\cite{9145807}: The paper proposes a novel approach to enhance temporal reasoning in Audio Question Answering (AQA) by introducing the Diagnostic Audio Question Answering (DAQA) dataset, which comprises natural sound events and programmatically generated questions to probe temporal reasoning skills. While adapting existing visual question answering methods to AQA reveals their limitations in temporal reasoning, the authors develop Multiple Auxiliary Controllers for Linear Modulation (MALiMo) to extend Feature-wise Linear Modulation (FiLM) by incorporating an additional auxiliary controller to process subsampled audio features, thereby enabling dynamic modulation of convolutional network processing based on both principal and supplementary inputs. This integrated approach addresses the challenge of in-depth temporal reasoning by facilitating relational and temporal analysis, leading to improved performance on DAQA without relying on spatial reasoning or static inputs.}
\item{\textsc{GPT-4o}~\cite{hurst2024gpt}: The paper proposes GPT-4o, an autoregressive omni model designed to process any combination of text, audio, image, and video inputs while generating text, audio, or image outputs through end-to-end training across modalities. By integrating Web Data, Code and Math, and Multimodal Data during pre-training, the model learns diverse reasoning skills and multimodal interpretation, while post-training alignment and red-teaming mitigate risks such as bias and harmful content. This unified approach enhances real-time responsiveness, multilingual performance, and multimodal understanding while addressing safety concerns through layered mitigations and external evaluations.}
\item{\textsc{GRU}~\cite{7410636}: The paper proposes a free-form, open-ended Visual Question Answering (VQA) task to generate natural language answers from images and questions, mirroring real-world scenarios like assisting the visually impaired. The approach leverages a large dataset (0.25M images, 0.76M questions, 10M answers) combining real images from MS COCO and abstract scenes to enable both low-level vision and high-level reasoning. By supporting diverse question types (e.g., fine-grained recognition, commonsense reasoning) and offering automatic evaluation through open-ended or multiple-choice formats, the framework addresses the need for detailed image understanding and multi-modal knowledge integration, advancing AI-complete challenges beyond generic captioning.}
\item{\textsc{HCAttn}~\cite{10.5555/3157096.3157129}: The paper proposes a hierarchical co-attention model for Visual Question Answering (VQA) that jointly reasons about image and question attention to improve answer accuracy. It introduces a co-attention mechanism to simultaneously perform question-guided visual attention (to identify relevant image regions) and image-guided question attention (to focus on key words), while employing a hierarchical question representation through word-level embeddings, phrase-level 1D CNNs (to capture n-gram features), and question-level LSTMs (to encode contextual meaning). By recursively combining co-attended features across these levels, the model addresses challenges like linguistic variation and multi-modal alignment, enhancing robustness and fine-grained understanding for VQA tasks.}
\item{\textsc{HCRN}~\cite{le2020hierarchical}: The paper proposes a general-purpose neural unit for video question answering that enables hierarchical relational reasoning and multimodal fusion. The Conditional Relation Network (CRN) processes input object arrays through sparse high-order relations while modulating encodings with contextual features, allowing flexible replication and stacking into Hierarchical CRNs (HCRN). The architecture integrates appearance features with clip motion as initial context, then progressively incorporates linguistic context and video-level motion through layered CRNs to enable multi-step reasoning. By hierarchically combining localized clip relations with global video and question contexts, HCRN addresses challenges of modeling distant temporal dependencies and heterogeneous modalities in VideoQA, demonstrating robust performance across diverse question types requiring appearance, motion, and temporal reasoning.}
\item{\textsc{HME}~\cite{fan2019heterogeneous}: The paper proposes a novel VideoQA framework that integrates heterogeneous memory and multimodal attention to enhance video-question reasoning. It introduces a heterogeneous memory module to jointly learn global context from appearance and motion features through synchronized attention, while a redesigned question memory captures complex semantics and highlights queried subjects by storing global contexts. These components interact through a multimodal fusion layer that aligns visual and textual hints via self-updated attention, enabling multi-step reasoning. By unifying feature integration with attention learning and maintaining global context throughout, the approach addresses challenges of spatiotemporal alignment and complex question semantics, improving VideoQA performance without separating feature and attention steps.}
\item{\textsc{LAST-Att}~\cite{10484352}: The paper proposes a method to address data bias in audio-visual question answering (AVQA) by constructing a balanced dataset and introducing an enhanced multimodal model. It identifies skewed answer distributions in the MUSIC-AVQA dataset and rectifies them by collecting complementary videos and questions to ensure uniform answer spread, particularly for binary questions, resulting in the MUSIC-AVQA v2.0 benchmark. The baseline model strengthens audio-visual-text interrelations through a pretrained Audio-Spectrogram-Transformer (AST) branch for audio grounding and cross-modal pixel-wise attention to align audio and visual spatial maps. By integrating these components, the approach mitigates modality neglect and improves reasoning across vision, audio, and language, establishing a robust foundation for unbiased AVQA evaluation.}
\item{\textsc{LAViT}~\cite{yun2021pano}: The paper proposes a novel benchmark for grounded audio-visual question answering on 360° videos to address spherical spatial reasoning and audio-visual relationships. It introduces the Pano-AVQA dataset with 51.7K QA pairs, featuring bounding-box grounding for two task types: spherical spatial relation QAs to assess relative object positioning on a sphere, and audio-visual relation QAs to link sounds with visual sources. Through quaternion-based spatial embeddings and multimodal training objectives, the framework integrates panoramic audio-visual cues while addressing challenges like spherical distortion and diverse sound localization. This holistic approach enhances semantic understanding of omnidirectional environments without relying on predefined fields of view.}
\item{\textsc{LAVisH}~\cite{lin2023vision}: The paper proposes adapting frozen vision transformers (ViTs) pretrained on visual data to audio-visual tasks without finetuning their original parameters. This is achieved through a latent audio-visual hybrid (LAVISH) adapter, which injects trainable parameters into each ViT layer to enable audio specialization and cross-modal fusion. The LAVISH adapter employs latent tokens to compress modality-specific information, reducing the quadratic cost of standard cross-attention while facilitating bidirectional audio-visual interaction. By integrating these components, the approach addresses the inefficiency of modality-specific models and costly audio pretraining, enabling frozen ViTs to leverage shared representations for enhanced audio-visual understanding without external encoders or extensive parameter updates.}
\item{\textsc{LSTTA}~\cite{liu2023parameter}: The paper proposes a parameter-efficient transfer learning approach for audio-visual-language tasks by introducing the Long Short-Term Trimodal Adapter (LSTTA), which integrates pre-trained unimodal/bimodal models without full fine-tuning. LSTTA employs a long-term semantic filtering module to suppress redundant video frames by characterizing temporal importance, while the short-term semantic interaction module models local cross-modal alignments through two sub-modules (AL2V and VL2A) to facilitate fine-grained information transfer. By combining these complementary mechanisms, LSTTA addresses the challenges of uneven global semantics and unannotated local correspondences in trimodal learning, enhancing performance on tasks like Music-AVQA and CMU-MOSEI without requiring large-scale trimodal pretraining.}
\item{\textsc{MAVEN}~\cite{ma2025fortisavqa}: The paper proposes a robust multimodal reasoning framework for Audio-Visual Question Answering (AVQA) to address dataset biases and enhance model robustness. It introduces FortisAVQA, a novel dataset constructed by rephrasing test questions to diversify linguistic forms and introducing distribution shifts to evaluate performance across frequent and rare question types. The Multimodal Audio-Visual Epistemic Network (MAVEN) employs a Multifaceted Cycle Collaborative Debias (MCCD) strategy to mitigate bias learning by enlarging distribution differences between unimodal and multimodal logits through KL divergence optimization while using cycle guidance to align unimodal logit distributions. This integrated approach reduces reliance on spurious correlations in individual modalities, improving generalization across in-distribution and out-of-distribution scenarios without requiring balanced training data.}
\item{\textsc{MCAN}~\cite{8953581}: The paper proposes a deep Modular Co-Attention Network (MCAN) to enhance visual question answering (VQA) by jointly modeling intra- and inter-modal interactions through a modular architecture. The framework integrates Self-Attention (SA) units to capture dense word-to-word and region-to-region relationships within questions and images, while Guided-Attention (GA) units model word-to-region cross-modal dependencies. By cascading Modular Co-Attention (MCA) layers composed of SA and GA units, MCAN enables deep reasoning while addressing the limitations of shallow co-attention models. This integrated approach improves fine-grained semantic understanding by simultaneously refining self-attention within modalities and guided-attention across modalities, leading to more accurate visual-textual alignment and robust performance on complex VQA tasks.}
\item{\textsc{MCCD}~\cite{ma2024look}: The paper proposes a robust framework for Audio-Visual Question Answering (AVQA) to address dataset biases and enhance model robustness. It introduces MUSIC-AVQA-R, a novel dataset crafted by rephrasing test questions and introducing distribution shifts to evaluate performance on both frequent and rare samples, while the Multifaceted Cycle Collaborative Debiasing (MCCD) strategy mitigates bias learning by enlarging distribution differences between uni-modal and multi-modal logits and employing cycle guidance to align uni-modal distributions. This integrated approach ensures diverse question evaluation and reduces bias dependency, improving generalization across in- and out-of-distribution scenarios without relying on balanced training data.}
\item{\textsc{Meerkat}~\cite{10.1007/978-3-031-73039-9_4}: The paper proposes an audio-visual LLM for fine-grained spatio-temporal grounding in images and audio, addressing the limitations of existing MLLMs in handling fine-grained tasks. It introduces a modality alignment module based on optimal transport to learn cross-modal patch alignment in a weakly-supervised manner, while a cross-attention module enforces audio-visual consistency to improve joint representation learning. These components are integrated through the AVFIT dataset (3M instruction samples) and MeerkatBench, a unified benchmark for five tasks, enabling the model to tackle challenges like disparate task formats and lack of large-scale training data. The approach enhances performance by unifying spatial and temporal grounding capabilities, achieving state-of-the-art results across diverse audio-visual tasks.}
\item{\textsc{OPM}~\cite{wei2024fly}: The paper proposes an adaptive modulation approach to address imbalanced multimodal learning by dynamically balancing uni-modal optimization during joint training. It introduces On-the-fly Prediction Modulation (OPM) to weaken dominant modality influence in the feed-forward stage by probabilistically dropping its features, while On-the-fly Gradient Modulation (OGM) mitigates gradient dominance in back-propagation through adaptive noise injection. By monitoring inter-modal discriminative discrepancies, these strategies jointly alleviate under-optimization of weaker modalities while preserving dominant modality contributions. The integrated framework enhances multimodal representation learning across diverse tasks by ensuring balanced feature optimization without additional training overhead, as validated through extensive experiments on audio-visual benchmarks.}
\item{\textsc{OneLLM}~\cite{han2024onellm}: The paper proposes a unified framework to align multiple modalities with language using a shared architecture, eliminating the need for modality-specific encoders. It introduces lightweight modality tokenizers to convert input signals into tokens, while a universal encoder (CLIP-ViT) extracts cross-modal features and a universal projection module (UPM) dynamically routes mixed projection experts to map diverse modalities into the LLM's embedding space. Through progressive alignment and a curated multimodal instruction dataset spanning eight modalities, the integrated approach overcomes scalability limitations of prior MLLMs by unifying encoding and projection, enabling flexible modality expansion and enhanced multimodal understanding without architectural redundancy.}
\item{\textsc{PSAC}~\cite{10.1609/aaai.v33i01.33018658}: The paper proposes a novel self-attention-based architecture for video question answering (VQA) to overcome the limitations of RNNs in modeling long-range dependencies and parallel processing. It introduces Positional Self-Attention (PSA) to capture global dependencies in video and question sequences by attending to all positions while incorporating absolute positional encodings to preserve temporal/spatial information. Through Video-based PSA (VPSA) and Question-based PSA (QPSA), the model encodes video frames and textual questions in parallel. A Video-Question Co-Attention (VQ-Co) block then simultaneously attends to relevant visual and textual features via bidirectional attention, enhancing cross-modal alignment. By integrating PSA with co-attention, the framework efficiently models complex video-question interactions, addressing challenges in sequential data processing and multimodal fusion while improving accuracy and computational efficiency.}
\item{\textsc{PSTP-Net}~\cite{li2023progressive}: The paper proposes a progressive spatio-temporal perception framework for audio-visual question answering (AVQA) to address challenges in complex multi-modal video understanding. The Temporal Segment Selection Module (TSSM) identifies relevant video segments to reduce redundancy, while the Spatial Region Selection Module (SRSM) locates question-aware visual patches within selected segments to enhance spatial reasoning. The Audio-guided Visual Attention Module (AVAM) models audio-visual associations by aligning sound features with visual patches. By progressively integrating these components, the approach effectively filters irrelevant content, localizes key spatio-temporal regions, and strengthens cross-modal interactions, leading to improved scene understanding and question answering performance.}
\item{\textsc{QaP}~\cite{liang2024querying}: The paper proposes a parameter-efficient multimodal language model learning strategy that bridges modalities through query-based prompts and lightweight resampling. The core innovation involves Querying Prompts (QP) to simultaneously extract modality information and interact with text, while Text-Conditioned Resamplers (TCR) adaptively inject text-relevant multimodal features into frozen language model layers. By integrating QP and TCR, the approach efficiently compresses modality inputs and leverages the model's inherent fusion capabilities, addressing computational inefficiency and redundancy in traditional projection-based methods while outperforming existing techniques across multiple multimodal tasks with minimal trainable parameters.}
\item{\textsc{Qwen2.5-VL}~\cite{bai2025qwen2}: The paper proposes Qwen2.5-VL, a vision-language model advancing multimodal understanding through enhanced visual recognition, object localization, and document parsing while addressing computational and contextual challenges. Key innovations include dynamic resolution processing to handle varying image sizes and video durations, absolute time encoding to improve temporal dynamics perception, and a native dynamic-resolution ViT with Window Attention to reduce overhead while preserving resolution. By integrating these components, the model achieves robust performance in fine-grained visual tasks, long-video comprehension, and real-world agentic applications without task-specific fine-tuning, while maintaining strong linguistic capabilities inherited from Qwen2.5 LLM. The approach overcomes bottlenecks in computational complexity and inconsistent sequence-length performance, enabling precise spatial-temporal reasoning and cross-domain generalization.}
\item{\textsc{RefAtomNet}~\cite{peng2024referring}: The paper proposes a novel approach for Referring Atomic Video Action Recognition (RAVAR) to identify atomic actions of a specific person guided by textual descriptions and video data, addressing limitations in traditional action recognition. Key components include RefAtomNet, which employs a multi-stream architecture connecting video, text, and location-semantic streams to interpret referring expressions and localize target individuals, while cross-stream agent attention and token fusion enhance relevance filtering across modalities. This integrated approach tackles challenges like irrelevant visual distractions and enables end-to-end action recognition for referred individuals, outperforming existing methods in RAVAR without requiring manual pre-processing. The RefAVA dataset with 36,630 annotated instances supports this task.}
\item{\textsc{VALOR}~\cite{10.1109/TPAMI.2024.3479776}: The paper proposes a Vision-Audio-Language Omni-Perception pretraining model (VALOR) to jointly model tri-modality interactions for understanding and generation tasks. It employs three single-modality encoders to process vision, audio, and language separately, while a multimodal decoder enables conditional text generation through two pretext tasks: Multimodal Grouping Alignment (MGA) projects modalities into a shared space to align vision-language, audio-language, and audiovisual-language groups via contrastive learning, and Multimodal Grouping Captioning (MGC) reconstructs masked text tokens conditioned on visual, auditory, or combined inputs to enhance generative capabilities. By integrating these components with a large-scale human-annotated dataset (VALOR-1M), the approach addresses the limitations of existing bimodal systems, enabling comprehensive cross-modal alignment and flexible text generation across diverse modality combinations for downstream tasks like retrieval, captioning, and question answering.}
\item{\textsc{VAST}~\cite{10.5555/3666122.3669307}: The paper proposes an omni-modality foundation model to enhance video-text cross-modality learning by integrating vision, audio, and subtitle information. It introduces VAST-27M, a large-scale dataset automatically generated through a two-stage pipeline: first training separate vision and audio captioners to produce single-modality descriptions, then employing an LLM to synthesize these with subtitles into omni-modality captions. The VAST model leverages three modality encoders and cross-attention-based text fusion, trained with objectives (OM-VCC/VCM/VCG) to unify multi-modal understanding. This approach addresses the lack of comprehensive video-text corpora by automating caption generation, enabling joint modeling of complementary modalities to improve performance on diverse downstream tasks like retrieval, captioning, and QA without manual annotation costs.}
\item{\textsc{VITA}~\cite{fu2024vita}: The paper proposes VITA, an open-source Multimodal Large Language Model (MLLM) capable of simultaneous processing and interactive analysis across video, image, text, and audio modalities. Starting with Mixtral 8×7B as a language foundation, it expands Chinese vocabulary through bilingual instruction tuning to enhance multilingual proficiency, while endowing visual and audio capabilities via two-stage multi-task learning for multimodal alignment and instruction tuning. To improve interaction, VITA introduces state tokens to distinguish input queries for non-awakening interaction and employs a duplex pipeline deployment scheme, where one model generates responses while another monitors environmental inputs, enabling audio interrupt interaction. This integrated approach addresses the lack of open-source models with unified multimodal processing and natural interaction, advancing seamless multimodal understanding and human-computer engagement without relying on wake-up words or sequential query handling.}
\item{\textsc{VideoLLaMA-2}~\cite{cheng2024videollama}: The paper proposes VideoLLaMA 2, a Video Large Language Model designed to enhance spatial-temporal modeling and audio understanding in multimodal video tasks. It introduces a Spatial-Temporal Convolution (STC) connector to capture intricate spatial and temporal dynamics in video data, while integrating an Audio Branch through joint training to incorporate audio cues for richer multimodal understanding. By combining these components, the model addresses challenges in processing temporal dynamics and audio-visual synchronization, improving performance in video question answering and captioning tasks without compromising contextual integrity or processing efficiency.}

\end{itemize}

\end{document}